\documentclass{article}

\pdfoutput=1

\usepackage{authblk}
\usepackage{epsfig}
\usepackage{epstopdf}
\usepackage{natbib}
\usepackage{fp}
\usepackage[capbesideposition=right]{floatrow}
\usepackage{wrapfig}
\usepackage[hmargin=0.75in,vmargin=1in]{geometry}

\newcommand{\Rs}{$ R_{\odot} $}
\def\ion[#1 #2]{#1\,{\sc #2}}

\begin{document}

\title{CME Induced Thermodynamic Changes in the Corona as Inferred from Fe XI and Fe XIV Emission Observations during the 2017 August 21 Total Solar Eclipse}

\author[1]{Benjamin Boe}

\author[1]{Shadia Habbal}

\author[2]{Miloslav Druckm\"uller}

\author[3,4]{Adalbert Ding}

\author[2]{Jana Hod\'erova}

\author[2]{Pavel \v Starha}

\affil[1]{\small Institute for Astronomy, University of Hawaii, Honolulu, HI 96822, USA}
\affil[2]{\small Faculty of Mechanical Engineering, Brno University of Technology, Technick 2, 616 69 Brno, Czech Republic}
\affil[3]{\small Institute of Optics and Atomic Physics, Technische Universitaet Berlin, Germany}
\affil[4]{\small Institut f\"ur Technische Physik, Berlin, Germany}

\maketitle

\begin{abstract}
We present the first remote sensing observations of the impact from a Coronal Mass Ejection (CME) on the thermodynamic properties of the solar corona between 1 and 3 \Rs. Measurements of the \ion[Fe xi] (789.2 nm) and \ion[Fe xiv] (530.3 nm) emission were acquired with identical narrow-bandpass imagers at three observing sites during the 2017 August 21 Total Solar Eclipse. Additional continuum imagers were used to observe K+F corona scattering, which is critical for the diagnostics presented here. The total distance between sites along the path of totality was 1400 km, corresponding to a difference of 28 minutes between the times of totality at the first and last site. These observations were used to measure the \ion[Fe xi] and \ion[Fe xiv] emission relative to continuum scattering, as well as the relative abundance of Fe$^{10+}$ and Fe$^{13+}$ from the line ratio. The electron temperature ($T_e$) was then computed via theoretical ionization abundance values. We find that the range of $T_e$ is $1.1-1.2 \times 10^6$ K in coronal holes and $1.2-1.4 \times 10^6$ K in streamers. Statistically significant changes of $T_e$ occurred throughout much of the corona between the sites as a result of serendipitous CME activity prior to the eclipse. These results underscore the unique advantage of multi-site and multi-wavelength total solar eclipse observations for probing the dynamic and thermodynamic properties of the corona over an uninterrupted distance range from 1 to 3 \Rs. 

\end{abstract} 
%\keywords{Sun: corona, eclipses, Sun: coronal mass ejections} 

\section{Introduction} 
\label{intro}
First indications of the high electron temperature, $T_e$, of the solar corona came from a previously unknown emission line observed during the 1869 Total Solar Eclipse (TSE) by \cite{Young1870, Young1871, Young1872}. This emission line was subsequently identified as \ion[Fe xiv] (530.3 nm) by \cite{Grotrian1934, Grotrian1939} and \cite{Edlen1943}. It took over six decades to recognize that the \ion[Fe xiv] spectral line came from a ground state magnetic dipole transition of Fe$^{13+}$, whose presence implied that the coronal plasma has a $T_e > 10^6$ K. Shortly thereafter, theories emerged to explain the cause of the high $T_e$ (e.g. magnetic wave heating; \citealt{Alfven1947}), and to predict the existence of the solar wind as a consequence of the corona's high temperature \citep{Parker1958}. Theoretical analyses have advanced a great deal since these early models, yet they are still largely unconstrained by observational data in the corona between 1 and 3 \Rs. Since $T_e$ is one of the main parameters which coronal heating and solar wind models must attempt to reproduce, it is important to have a complete understanding of its distribution in the corona.
\par

With the advent of space-based observations, inferences of the coronal $T_e$ have vastly expanded via remote sensing at wavelengths previously inaccessible from the ground, and by \textit{in situ} particle detector measurements of the solar wind. Ultraviolet and Extreme Ultraviolet (EUV) spectroscopy can be used to infer $T_e$ in the chromosphere and low corona ($< 1.5$ \Rs) from remote sensing observations of ionic emission lines (e.g. \citealt{Habbal1993, Raymond1997, Morgan2017}). Coronal $T_e$'s inferred from these studies typically range from $1$ to $ 4 \times 10^6$ K. Particle detectors \textit{in situ}, such as SWICS on Ulysses and ACE, have inferred coronal $T_e$ via measurements of charge to mass ratios of solar wind plasma \citep{Gloeckler1992, Gloeckler1998}. Ion abundance ratios are determined from charge to mass ratios which are then modeled to infer to $T_e$ in the corona at the freeze-in distance (e.g. \citealt{Hundhausen1968, Owocki1983b, Ko1997, Zurbuchen2002}). The \textit{in situ} measurements typically yield values between 1 and 3 $\times 10^6$ K \citep{Habbal2010a}, which vary spatially between structures such as coronal holes and streamers, as well as with transient events such as the passage of Coronal Mass Ejections (CMEs) \citep{Smith2003}.

\par
While highly valuable, inferences of $T_e$ based on \textit{in situ} and EUV measurements have some key limitations. \textit{In situ} analyses face difficulties in robustly tracing the origin of the ions back to the corona (e.g. \citealt{Galvin1997}), and EUV intensity drops very quickly with distance in the corona due to its dependence on the density squared (i.e. $n_e^2$; from collisional excitation, see Section \ref{time}). Furthermore, EUV imaging observations such as those provided by SDO/AIA have to contend with line crowding in the filter bandpasses, causing a highly complex temperature response function (e.g. \citealt{ODwyer2010, Boerner2012}). An iterative method is often used to settle on a best fit temperature, although this technique can produce different results with the same data depending on the exact iteration method used \citep{Wit2013}. This technique also assumes a state of ionization equilibrium in the corona, which is not necessarily valid beyond $\approx $ 1.2 \Rs, as demonstrated by \cite{Landi2012b} through modeling and by \cite{Boe2018} from inferences of Fe$^{10+}$ (\ion[Fe xi], 789.2 nm) and Fe$^{13+}$ (\ion[Fe xiv], 530.3 nm) freeze-in distances which can be as low as 1.2 \Rs \ in coronal holes.
\par
In this work we use imaging observations of \ion[Fe xi] and \ion[Fe xiv] emission taken from three sites during the 2017 August 21 Total Solar Eclipse as described in Section \ref{Data}. The data spanned almost 30 minutes of totality, enabling the inference of the spatial distribution and temporal variation of the \ion[Fe xi] and \ion[Fe xiv] line to continuum intensity ratios (see Section \ref{lineCont}), as well as the ionic density ratio and $T_e$ in the corona using the technique described in Section \ref{methods}. Spatial differences between coronal holes and streamers are presented in Section \ref{tempDist}. An analysis of the temporal variability of coronal thermodynamical properties between observing sites and a discussion on the effect that a Coronal Mass Ejection had on these observations are given in Section \ref{time}. Concluding remarks are given in Section \ref{conc}.
  
\section{Data}
\label{Data}
\subsection{Eclipse Observations}
\label{eclipse}
The imaging data of the solar corona used here were acquired during the 2017 August 21 Total Solar Eclipse at three separate sites spread over 1400 km along the path of totality in the United States. Mitchell, Oregon, was the first site to experience totality, followed by Mackay, Idaho, and finally Alliance, Nebraska, which viewed totality approximately half an hour after Mitchell. All three sites had cloudless skies and good seeing throughout totality. Details of the observing locations and eclipse conditions for each site can be found in Table \ref{table1}. 

\begin{table}[h]
\begin{center}
\begin{tabular}{ | c  | c | c | c | }
\hline
Observing Site & Mitchell, OR & Mackay, ID & Alliance, NE \\
\hline
Latitude & N 44$^o$ 31' 32.34" & N 44$^o$ 3' 12.59" & N 42$^o$ 5' 24.30" \\
\hline
Longitude & W 119$^o$ 54' 27.54" & W 109$^o$ 36' 39.67" & W 103$^o$ 0' 29.98" \\
\hline
Elevation (m) & 1100 & 1946 & 1300  \\
\hline
Solar Altitude (degrees) & 42.5-42.8 & 47.6-47.9 & 56.5-56.8 \\
\hline
C2 Time (UT) & 17:21:11.7 & 17:29:53.0 & 17:48:57.5 \\
\hline
C3 Time (UT) & 17:23:14.1 & 17:32:08.0 & 17:51:28 \\ 
\hline
Duration (minutes) & 2.04 & 2.25 & 2.51 \\
\hline
\end{tabular}
\caption{Observing locations and corresponding eclipse parameters}
\end{center}
\label{table1}
\end{table}

\par

Each observing site was equipped with a suite of narrowband imaging telescopes to observe optical line and continuum emission in the corona. All imagers had a 70 mm diameter aperture, a 300 mm focal length achromat lens and a 0.5 nm Fabry-P\'erot bandpass filter manufactured by Andover corporation. The emission lines of Fe$^{10+}$ (\ion[Fe xi], 789.2 nm) and Fe$^{13+}$ (\ion[Fe xiv], 530.3 nm) were observed using filtered imaging systems centered on the wavelength of line emission, which we will refer to as `on-band' hereafter. Additional continuum, or `off-band', imagers were used to measure the continuum emission at a nearby wavelength to the line emission for both \ion[Fe xi] and \ion[Fe xiv]  (see Section \ref{Calib}). Throughout the rest of this article we will refer to the roman numeral version (i.e. \ion[Fe xi] ) when discussing the observed line emission, and the ionic version (i.e. Fe$^{10+}$) when referring to the ions themselves. 
\par

While coronal emission lines provide the best metric for inferring $T_e$ in the corona (see Section \ref{methods}), the most dominant form of coronal emission at visible wavelengths is actually continuum radiation from Thompson scattering of solar photons by electrons, known as the K corona, combined with Mie scattering by dust particles, called the F corona or inner zodiacal light \citep{vandeHulst1950}. The off-band images can be treated as a measurement of the K+F corona at a well defined wavelength, while the on-band images capture the exact same K+F corona, albeit with the addition of forbidden ion emission lines (or E corona). These continuum observations are essential for measuring and differentiating between K+F scattering and spectral line emission (see Section \ref{lineCont}). Indeed, a single imaging observation at visible or infrared wavelengths is incapable of differentiating line emission from continuum radiation. Such an observation will not provide any physically meaningful information about the corona, other than morphological structure, unless the continuum emission is also observed and subtracted. 

\par

\begin{figure*}[h!]
\centering
\includegraphics[width = 6in]{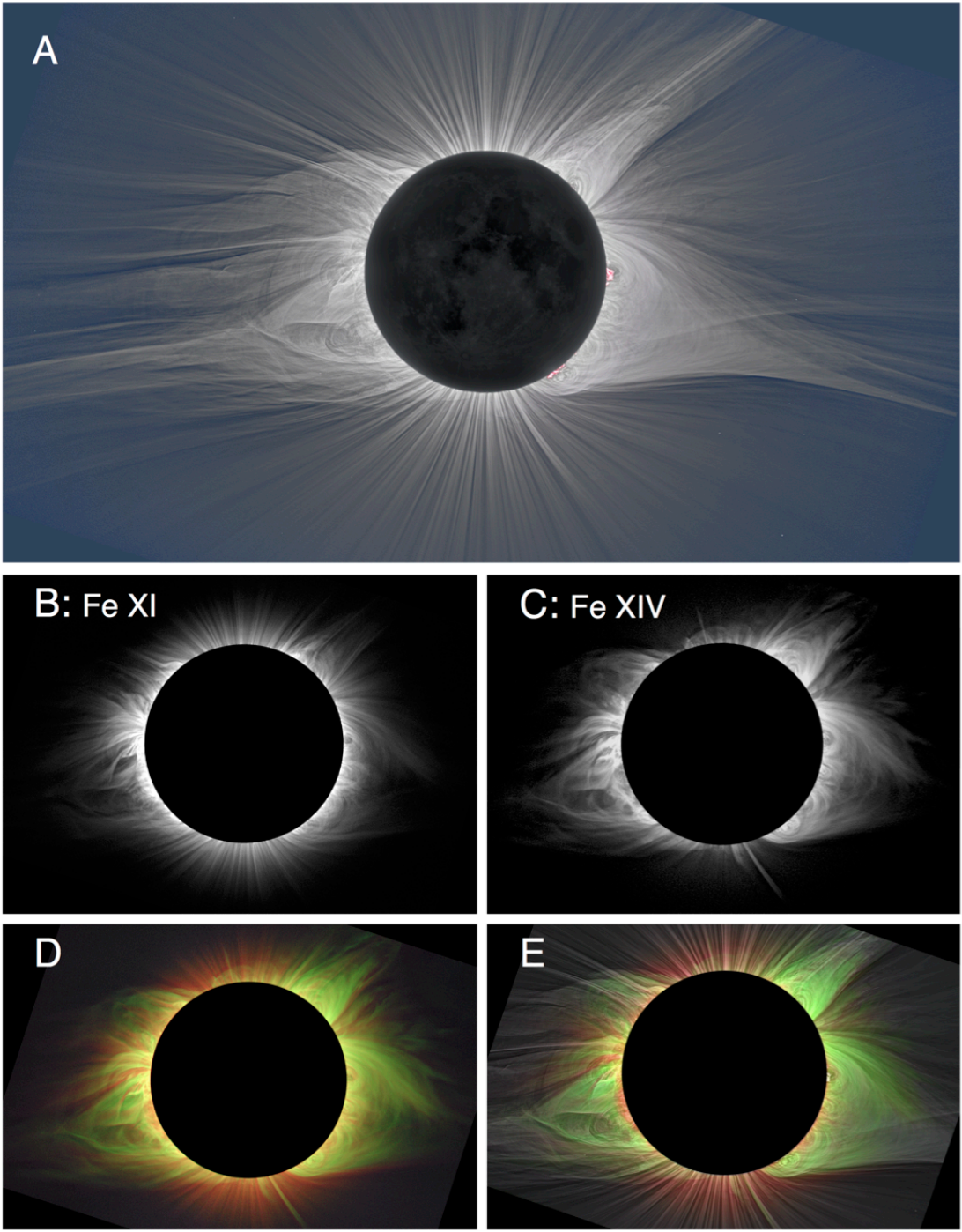}
\caption{Total solar eclipse observations from Mitchell, OR, on 2017 August 21. Solar north is vertically up in all images. A: Processed white light image of the solar corona. B: Processed image of the \ion[Fe xi] 789.2 nm emission at $\approx 1.1 \times 10^6$ K. C: Processed image of the \ion[Fe xiv] 530.3 nm emission at $\approx1.8 \times 10^6$ K. D: Composite of B (red) and C (green). E: Same as D overlaid on the white light image (Note that the corresponding continuum has been subtracted from the on band \ion[Fe xi] and \ion[Fe xiv] images prior to processing to produce the images shown here).}
\label{fig1}
\end{figure*}

An example of data collected from Mitchell, OR, is shown in Figure \ref{fig1}, including a broadband white light image which records the finely detailed structures in the corona (Fig \ref{fig1} A). Figures \ref{fig1} B and C then show the off-band subtracted \ion[Fe xi] and \ion[Fe xiv] emission (see Section \ref{Calib}), followed by two composite images. The data presented in this Figure were processed using the Adaptive Circular High-pass Filter (ACHF) method to enhance structural features in the corona, so they are not representative of the absolute coronal brightness. The ACHF method, introduced by \cite{Druckmuller2006}, has been used previously to process white light eclipse images (e.g. \citealt{Habbal2010a, Habbal2014, Alzate2017, Boe2018}). Note that the white light image (i.e. Fig \ref{fig1} A) is not used in our analysis except to provide context for the \ion[Fe xi] and \ion[Fe xiv] photometric observations.

%FIGURE 1
\par

\par
The 0.5 nm Fabry-P\'erot bandpass filters were mounted at the front of each telescope's objective lens in order to avoid angle dependent changes to the wavelength transmission. Our total field of view is within 1 degree of Sun center, which corresponds to a change in wavelength transmission of about 0.1 nm based on the equation provided by the manufacturer:
\begin{eqnarray}
\lambda = \lambda_0 \left(1- \left(\frac{n_m}{n_f} \right)^2 \sin^2{ \theta} \right)^{0.5},
\end{eqnarray}
where $n_m$ is the refractive index of the external medium (air) and $n_f$ is the refractive index of the filter. The wavelength shift induced by the filter with angular distance in the data is thus substantially smaller than the size of the transmission width and will not have an impact on our final results. There will also be a shift in wavelength due to the temperature of the filter, so each filter was equipped with an electric heater that maintains the temperature at a precision of 0.1 K to ensure the correct observational wavelength. 
\par
All of the Fabry-P\'erot filters were designed with full width at half maximum (FWHM) values of 0.5 nm, but there was some small variation between the filters. We performed a theoretical test on the filter profiles by modeling the line and continuum intensity that would be measured by the exact filter profiles compared to each other, after a solar disk calibration (see Section \ref{Calib}). We find that all filters can recover the same on- and off-band intensity to $< 1 \%$ for a synthetic solar continuum combined with a 0.2 nm Gaussian line emission profile. 

\par
The \ion[Fe xiv] systems had Atik 314L cameras, while Atik 414EX cameras were used with the \ion[Fe xi] systems due to their higher quantum efficiency in the near infrared. Both camera models have an array of 1400 x 1024 pixels that are each 6.5 $\mu$m in size, resulting in an angular resolution of about 4" per pixel. For this work we rebinned the original data in 2x2 pixel squares, hence increasing the signal to noise ratio at the cost of reducing spatial resolution to $\approx 8$''. Photometric uncertainties were computed using $\sigma = \sqrt{N}$, where $N$ is the number of counts in the raw data for a given pixel. The uncertainty of any pixel intensity ratio presented in this work (e.g. Sections \ref{lineCont} and \ref{temp}), was determined by propagating the individual photometric uncertainties using standard error propagation equations (e.g. for subtracted line emission, the uncertainty is the propagation of error from both the on band and continuum intensity measurements).
\par
During totality, the cameras cycled through a sequence of seven different exposure times from 0.1 to 6.4 seconds, spaced by factors of two in time. A range of exposure times is required to measure the $> 3$ orders of magnitude change of the coronal intensity between a heliocentric distance of 1 and 3 \Rs. Each exposure was individually reduced using traditional dark subtraction and flat fielding techniques. The sequence of observations were then aligned and stacked separately for each telescope system using a phase correlation method described in \cite{Druckmuller2009}. The sky brightness in each data set was measured as the average surface brightness below 0.05 \Rs \ (in front of the Moon) and subtracted from the data. 

\subsection{Data Calibration}
\label{Calib}

Relative calibrations between the telescope systems at each site were made using observations of the full solar disk after the partial phases of the eclipse had ended (after C4), through an additional Neutral Density (ND) filter mounted on the telescope aperture (to reduce the intensity of the solar disk). All ND filters used were designed to attenuate the intensity to $1.5 \times 10^{-5}$ of the incident intensity in the vicinity of the observed wavelength, where the same filter was used for each on- and off-band pair at given site. The entire intensity of the disk was integrated in the images (with the same reduction process used for the eclipse data) and used to scale the photometric response of the on- and off-band  instrument pairs at each site. The size of the solar disk in the image is around 5 $\times 10^5$ pixels, with each pixel having an intensity $>10^3$ counts, leading to a precise relative photometric calibration. The solar disk calibration enables a measurement of the line to continuum intensity ratio for \ion[Fe xi] and \ion[Fe xiv] at each site (see Section \ref{lineCont}).

\par
At Mitchell, Oregon, the continuum imagers had a central filter wavelength shifted by approximately 3 nm toward the blue relative to the emission line observations. These separations represent a Doppler velocity shift of 1200 and 1640 km s$^{-1}$ for \ion[Fe xi] and \ion[Fe xiv] respectively, which is significantly higher than the speeds expected in the corona or solar wind (e.g. \citealt{Smith2003, Habbal2010a}). Consequently, it can be safely assumed that these continuum observations are not contaminated by Doppler shifted line emission. At Mackay and Alliance however, the filters were shifted by only 0.94 and 1.4 nm for \ion[Fe xi] and \ion[Fe xiv], which correspond to a Doppler velocity of 530 km s$^{-1}$ for both ions. It is possible for coronal and solar wind plasma to be moving at this velocity (especially in the fast solar wind where Fe$^{10+}$ is very abundant), so Doppler shifted contamination is possible. In fact, we see strong evidence of Doppler shifted emission, especially in the \ion[Fe xi] continuum data (see Appendix \ref{DopplerShift}).

\par
Fortunately, at the Alliance site there was an additional continuum imager meant for the \ion[Ar X] (553.3 nm) emission line. The \ion[Ar X] line intensity was too faint for us to use in this work, but the \ion[Ar X] continuum observation at 552.3 nm is only separated by 22 nm ($> 12 000$ km s $^{-1}$) from the \ion[Fe xiv] line and so is usable as a replacement for the original \ion[Fe xiv] continuum observation. The 552.3 nm continuum observation also had solar disk calibration observations made with the same ND filter as used for the \ion[Fe xiv] calibration, so we used the nominal procedure to calibrate this substitute data. No additional continuum data was available at the Mackay site, so we used the unaltered \ion[Fe xiv] continuum data. Based on the Doppler emission seen in the Alliance \ion[Fe xiv] continuum data (after having the \ion[Ar X] continuum subtracted, see Appendix \ref{DopplerShift}), there is an average of about $5.7 \%$ Doppler contamination, with a maximum of about $12 \%$. To account for this possible Doppler bias, we introduced an additional $5.7 \%$ uncertainty to the error analysis of the Mackay data. For both Mackay and Alliance, the solar disk calibrations are still valid for producing a pristine line to continuum ratio metric for \ion[Fe xiv] (see Section \ref{lineCont}).
\par

The \ion[Fe xi] data at Mackay and Alliance are more complicated to calibrate given the Doppler shifted contamination. Unlike the relatively small impact of Doppler shifted \ion[Fe xiv], the \ion[Fe xi] continuum data at Mackay and Alliance show a strong contribution from Doppler shifted emission. Based on this contamination, we were forced to model the \ion[Fe xi] continuum data for Mackay and Alliance using the \ion[Fe xiv] and \ion[Ar x] continuum data respectively. An additional complication for creating a realistic \ion[Fe xi] continuum is possible reddening effects from the F corona. The K corona will have a neutral color, as Thompson scattering has no wavelength dependence whatsoever, whereas the F corona can be reddened at visible wavelengths due to dust diffraction scattering \citep{vandeHulst1947}. The K corona is the strongest component of the continuum below at least 1.3 \Rs \ everywhere in the corona throughout the solar cycle (e.g. \citealt{Koutchmy1985}, \citealt{Hiroshi1998} and ref. therein), and so the continuum in the corona will \textit{not} be substantially impacted by F corona reddening at 1.1 \Rs \citep{Roeser1978}. We used this fact to calibrate between the \ion[Fe xi] and \ion[Fe xiv] continuum data from Mitchell, where the off-band observations were free of Doppler contamination. The solar disk ND filter observations already calibrated each on- and off- band pair, but to calibrate the \ion[Fe xi] and \ion[Fe xiv] to each other, we took the average intensity ratio of the continuum data below 1.1 \Rs \  in a 30 degree region centered on the western streamer. In doing so, we created a map of the F corona red excess (see Appendix \ref{DopplerShift}), and provided the final step in the cross calibration between all datasets at Mitchell (i.e. $\frac{\rho(\nu_{k}) \ \epsilon_{k}}{\rho(\nu_{j}) \ \epsilon_{j}}$ in Equation \ref{eqnFinal}, Section \ref{temp}). 
\par

The \ion[Fe xi] continuum data is then modeled for Mackay and Alliance by taking the reddening correction for the F corona observed at Mitchell, and applying it to the calibrated \ion[Fe xiv] and \ion[Ar X] continuum data from Mackay and Alliance respectively. This technique provides the shape of the continuum for \ion[Fe xi] at Mackay and Alliance, but does not provide an intensity calibration. Given that the Mitchell data was pristine, we calibrated the line data (i.e. on - continuum) for Mackay and Alliance using the average line to continuum ratio found in the north coronal hole ($< 1.1$ \Rs) of the Mitchell data. This procedure is assuming that the \ion[Fe xi] line to continuum ratio in the coronal hole did not change over the 28 minutes between sites, which may not be the case. Nonetheless, this calibration provides a means for setting the lower limit on the size of changes during this time. If the coronal hole also changed, then the magnitude of changes (both $T_e$ and \ion[Fe xi] line to continuum ratio) presented in this work would only increase. Additionally, the F corona intensity is exceptionally constant throughout the solar cycle \citep{Morgan2007}, so any inferred changes between sites (see Sections \ref{lineCont} and \ref{time}) cannot be due to the F corona. 

\subsection{Spacecraft Observations}
\label{spacecraft}
\begin{figure*}[t!]
\centering
\includegraphics[width = 6in]{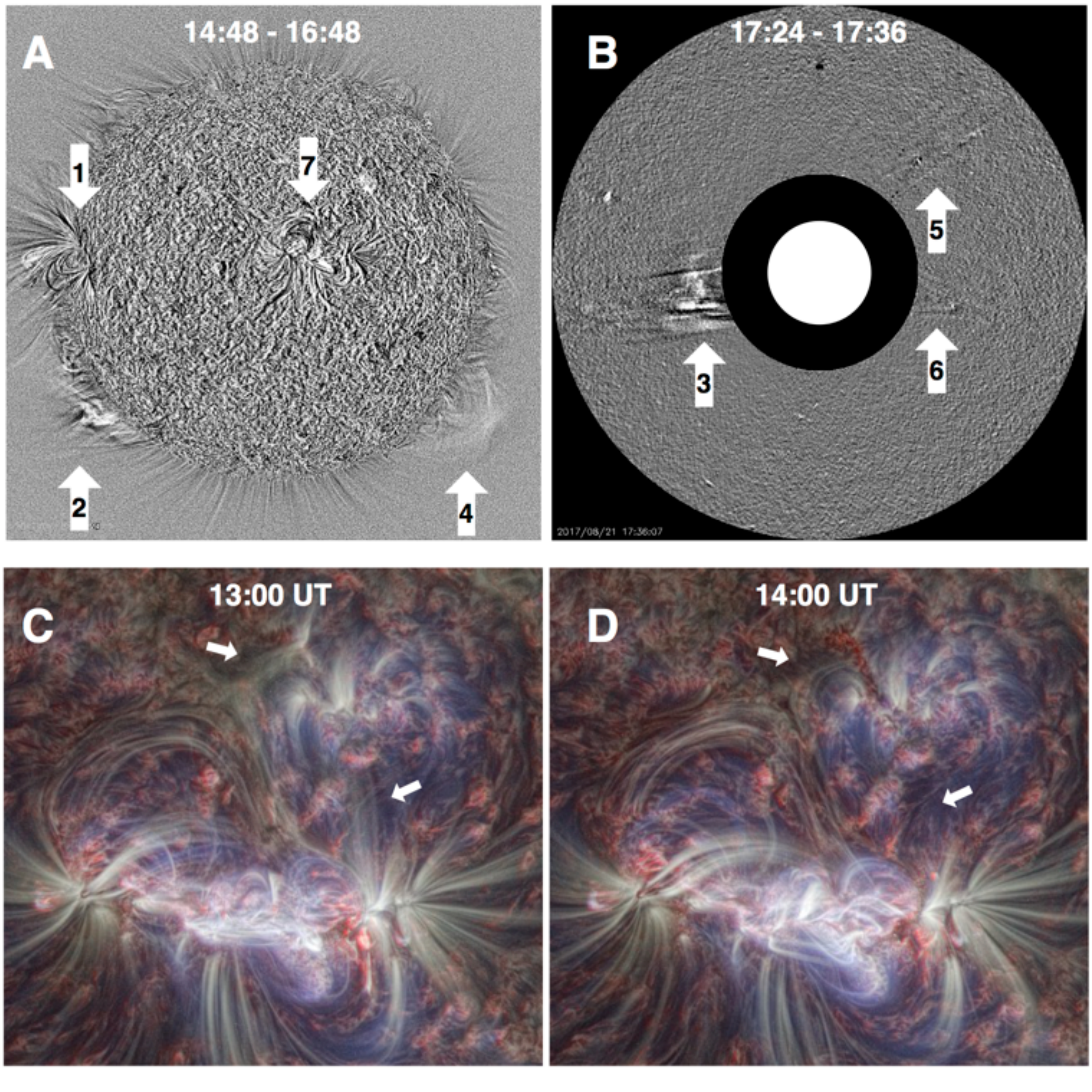}
\caption{A: Difference between two processed SDO/AIA 17.1 nm images showing changes over two hours prior to the eclipse (14:48 and 16:48 UT), with arrows indicating active region 12672 (arrow 1) and 12671 (arrow 7) along with some erupting structures that will impact the outer corona, namely a joint eruption of active region 12672 (arrow 1) and a prominence in the southeastern corona (arrow 2). B: Difference between two processed SOHO/LASCO-C2 images from 17:24 and 17:36 UT, showing eruptions (arrows 3, 5 and 6) in the corona occurring at the time of the eclipse. The inner white circle represents the approximate size of the solar disk. C and D: PM-NAFE processed AIA data of the disk center active region (12671) at 13:00 and 14:00 UT respectively. The arrows indicate changes in the PM-NAFE maps, likely due to a halo CME erupting from this active region (s ee Section \ref{spacecraft} for details on the image processing).} 
\label{fig2}
\end{figure*}

\begin{figure*}[t!]
\centering
\includegraphics[width = 6in]{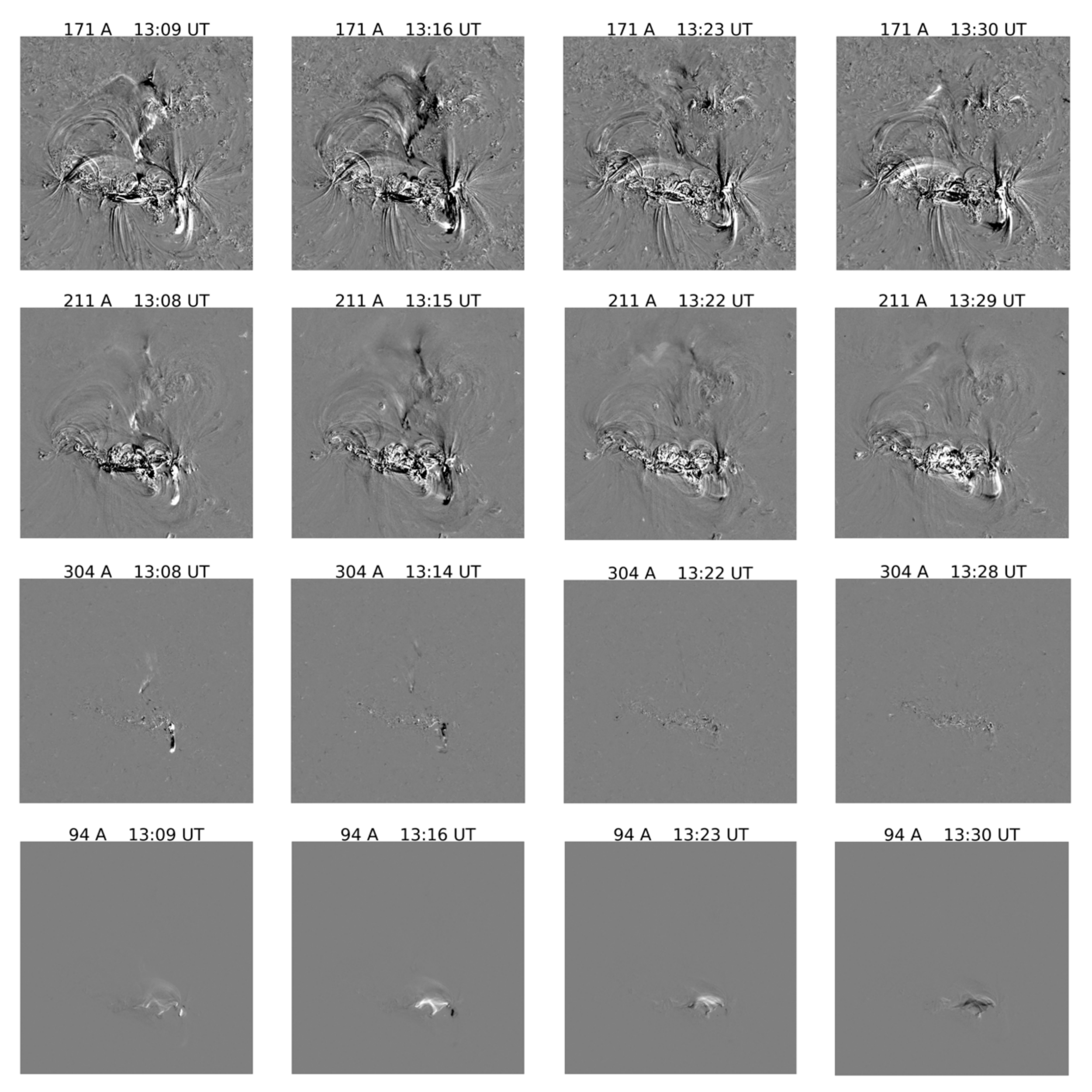}
\caption{Absolute difference images of SDO/AIA centered on active region 12671 (see arrow 7 in Fig \ref{fig2}), which was near disk center and 10 degrees north latitude at the time of the eclipse. Each frame shows the change over a $\approx 7$ minute period. From top to bottom the rows show emission changes from 17.1 nm (\ion[Fe ix]), 21.1 nm (\ion[Fe xiv]), 30.4 nm (\ion[He ii]), and 9.4 nm (\ion[Fe xviii]). These changes highlight a CME that occurred almost simultaneously with other eruptive events near the limb of the corona (see Figure \ref{fig2}).} 
\label{fig3}
\end{figure*}

\begin{figure*}[t!]
\centering
\includegraphics[width = 6in]{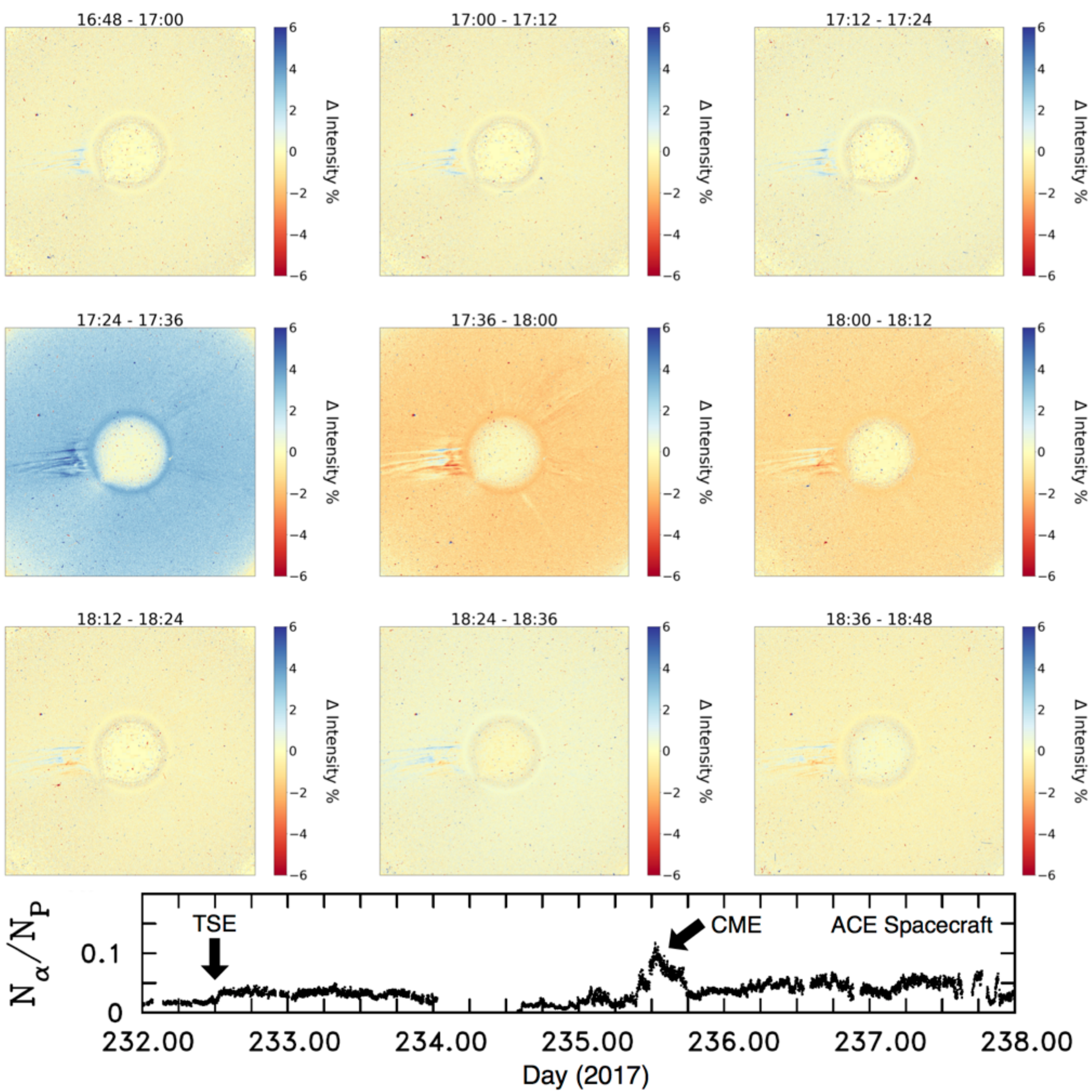}
\caption{Top: Percentage difference images of LASCO C2 coronagraph data indicating the total white light continuum intensity changes around the time of the eclipse. There is a substantial increase in emission ($> 3\%$) occurring between 17:24 and 17:36, which roughly coincide with the times of totality at Mitchell and Mackay. The increase, and subsequent decrease in emission, is evidence of halo-CME activity during the period of eclipse observations. Bottom: Alpha particle abundance relative to protons in the solar wind at L1 as observed by the ACE spacecraft. Arrows indicate the time of the total solar eclipse (TSE) and the signature of a CME arriving about 3 days after the eclipse. } 
\label{fig4}
\end{figure*}

A series of eruptive events occurred in the corona just prior to the time of the eclipse as illustrated by SDO/AIA and SOHO/LASCO spacecraft observations presented in Figures \ref{fig2}, \ref{fig3} and \ref{fig4}. Figure \ref{fig2} A shows the difference of two AIA 17.1 nm images processed using the Multi-scale Gaussian Normalization technique (MGN) from \cite{Morgan2014} just prior to the eclipse (at 14:48 and 16:48 UT). Panel B shows a differenced LASCO-C2 image (at 17:24 and 17:36 UT) that was processed using the Dynamic Separation Technique (DST) from \cite{Morgan2012, Morgan2015}. Panels C and D show PM-NAFE (Planckian Mapping-Noise Adaptive Fuzzy Equalization) processed AIA data of the disk center active region (AR 12671). PM-NAFE is an algorithm developed by \cite{Druckmuller2013} to combine SDO/AIA data into a single composite image, where the color of the image corresponds to the average electron temperature inferred by the relative flux of the AIA channels. Figure \ref{fig3} contains a series of difference images for the same disk center active region that were created by directly subtracting pairs of two AIA images, so they indicate changes in the absolute EUV emission within the bandpass. This technique is different from the panels in Figure \ref{fig2}, which were first processed with MGN, DST, or PM-NAFE to enhance structural features. Figure \ref{fig4} shows the time evolution of the continuum corona as seen by LASCO C2, including a transient spike in continuum emission which occurred around the time period between Mackay and Alliance. The bottom of Figure \ref{fig4} also contains \textit{in situ} alpha particle abundance data from the ACE spacecraft \citep{Stone1998} for several days after the eclipse. 

%FIGURE 2

%FIGURE 3

%FIGURE 4

\par
The most notable event recorded by the spacecraft observations prior to the eclipse was a CME that erupted from active region 12672 on the east limb (arrow 1 in Fig. \ref{fig2} A) coupled with a filament eruption to the south (arrow 2 in Fig. \ref{fig2} A). Around approximately 13:00 UT on the day of the eclipse, the CME and faint prominence core began moving slowly outward as seen in AIA images. By 16:00 UT, the CME had moved out of the AIA field of view ($\approx 1.2$ \Rs). The CME was first visible by LASCO C2 on SOHO ($\approx 2.0$ \Rs, see arrow 3 in Fig. \ref{fig2} B) by 17:00 UT until at least 22:00 UT (the first eclipse in our dataset began at 17:21). During the time interval of the eclipse observations at all three sites, the CME front had made its way toward the edge of our images ($\approx 2$ to $3$ \Rs), having disturbed the lower corona over the preceding few hours. There were two other small eruptive like regions in the north- and south-western corona (see arrows 4, 5 and 6 in Fig. \ref{fig2}) that are noticeable in the AIA 17.1 nm and LASCO C2 differenced images. These smaller eruptions do not necessarily generate a CME (i.e. an eruption that fully escapes from the Sun), but they can still have an impact on the thermodynamics in the corona (as shown in Sections \ref{lineCont}, \ref{temp}, and \ref{time}).  

\par
At the same time that the eruptions near the limb of the Sun began, there were motions occurring in active region 12671 (see C and D in Fig. \ref{fig2}, and \ref{fig3}) that appear to be a CME event near the center of the solar disk. Around 13:00 UT on the day of the eclipse, outward motions and the following disappearance of closed field lines on the north side of the active region are seen by the 17.1 (\ion[Fe ix]) and 21.1 nm (\ion[Fe xiv]) channels, while the chromosphere (30.4 nm, \ion[He ii]) and high temperature plasma (9.4 nm, \ion[Fe xviii]) show a short duration spike in flare-like emission that is characteristic of the base of an erupting CME from an active region (e.g. \citealt{Fletcher2011} and ref. therein). Given the location of the active region, the CME is likely moving almost directly toward the Earth, perhaps slightly northward given the latitude of the active region. Structural variation in the corona caused by such a relatively small halo-CME would be exceedingly difficult to detect in the LASCO C2 coronagraph data, but the continuum intensity variation is detectable. The LASCO C2 coronagraph data show a transient increase in white light emission occurring during the period of the eclipse (see Fig \ref{fig4}) which is likely caused by this halo-CME. The ACE spacecraft \citep{Stone1998} at the L1 Lagrange point also observed a transient spike in the alpha particle abundance (He abundance is an indicator of interplanetary CMEs; e.g. \citealt{Borrini1982}). The alpha particle abundance was initially stable around $\approx 3 \%$ of the proton density for several days, then the abundance rose to between 5 and 10 $\%$ for several hours about three days after the eclipse (see bottom of Fig \ref{fig4}). The time delay would correspond to $\approx 545$ km s$^{-1}$ assuming the CME traveled at a constant velocity after eruption at 13:00 UT on the day of the eclipse. The presence of a halo-CME is further supported by changes in Doppler emission of \ion[Fe xi] (see Appendix \ref{DopplerShift}). The simultaneity of these two CMEs and sub-eruptive events in the streamers hint that there was some magnetic connectivity between these regions leading to `sympathetic CME' behavior (i.e. \citealt{Moon2003}). It is likely that these events caused global scale perturbations in the corona (see Sections \ref{temp}, and \ref{time}).

\section{Observational Metrics}
\label{metrics}

\subsection{Line to Continuum Ratio}
\label{lineCont}
\begin{figure*}[t!]
\centering
\includegraphics[width = 6in]{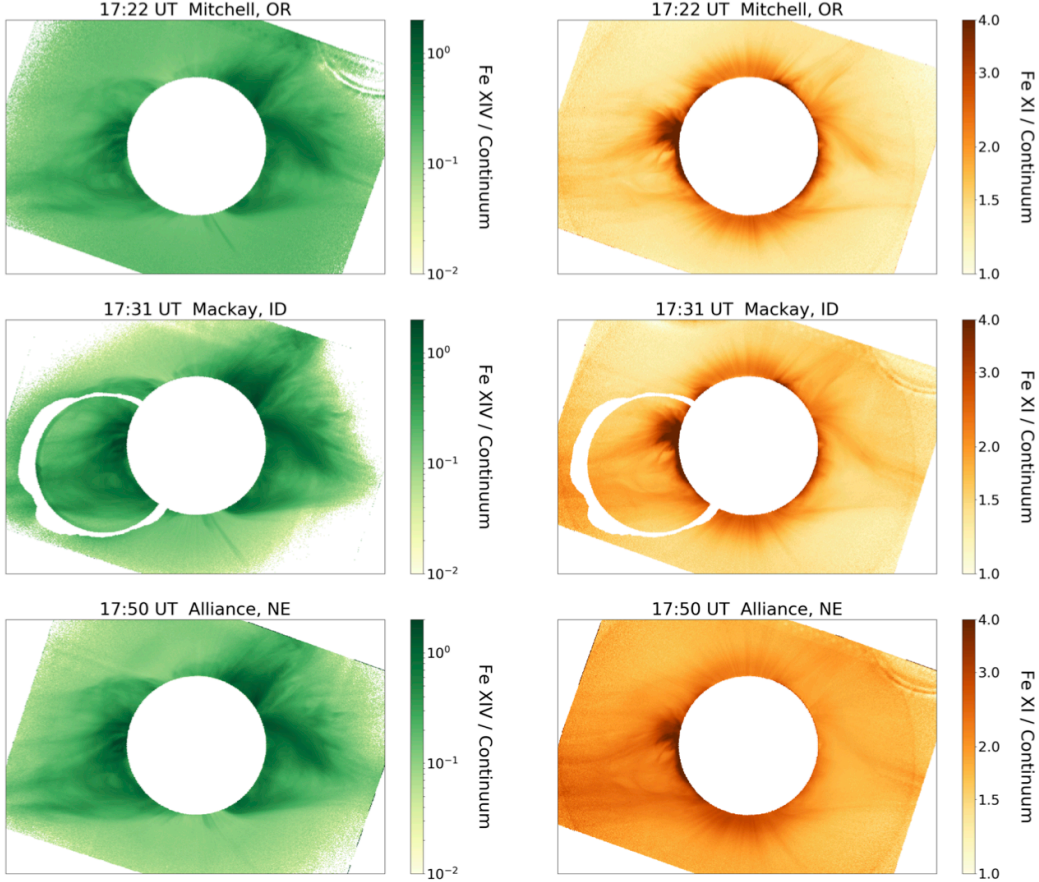}
\caption{Line to continuum ratio of \ion[Fe xiv] (left) and \ion[Fe xi] (right) made at three separate sites (see Section \ref{lineCont}). The line and continuum data were calibrated using the technique outlined in Section \ref{Calib}. The white ring in the middle panels indicates the location of a ghost image in the raw data produced by internal reflections in the optics (see Appendix \ref{ghost}).}
\label{fig5}
\end{figure*}

\begin{figure*}[t!]
\centering
\includegraphics[width = 6in]{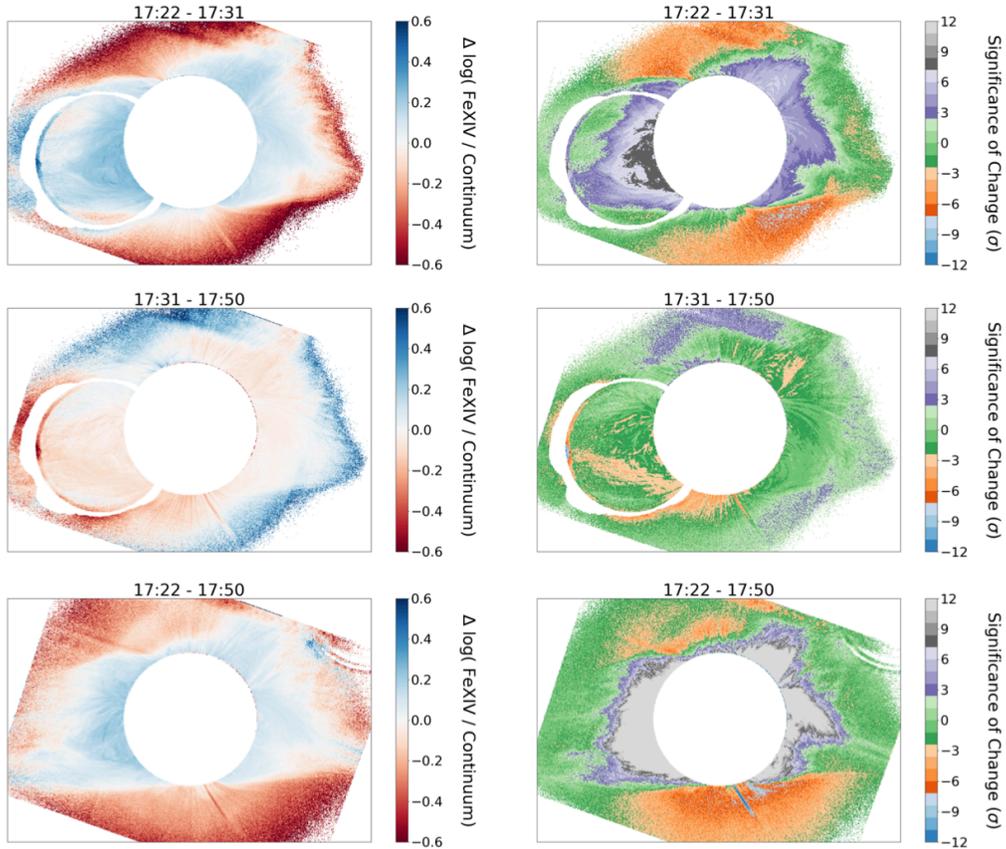}
\caption{Left: Change in the $Log_{10}$ of the \ion[Fe xiv] line to continuum ratio between the three observing sites (see Section \ref{lineCont} and \ref{time}). Right: Statistical significance of these changes between sites. Top: Mackay - Mitchell; Middle: Alliance - Mackay; Bottom: Alliance - Mitchell. The white rings in the top and middle panels are the same indication of a ghost image as in Figure \ref{fig5}.}
\label{fig6}
\end{figure*}

\begin{figure*}[t!]
\centering
\includegraphics[width = 6in]{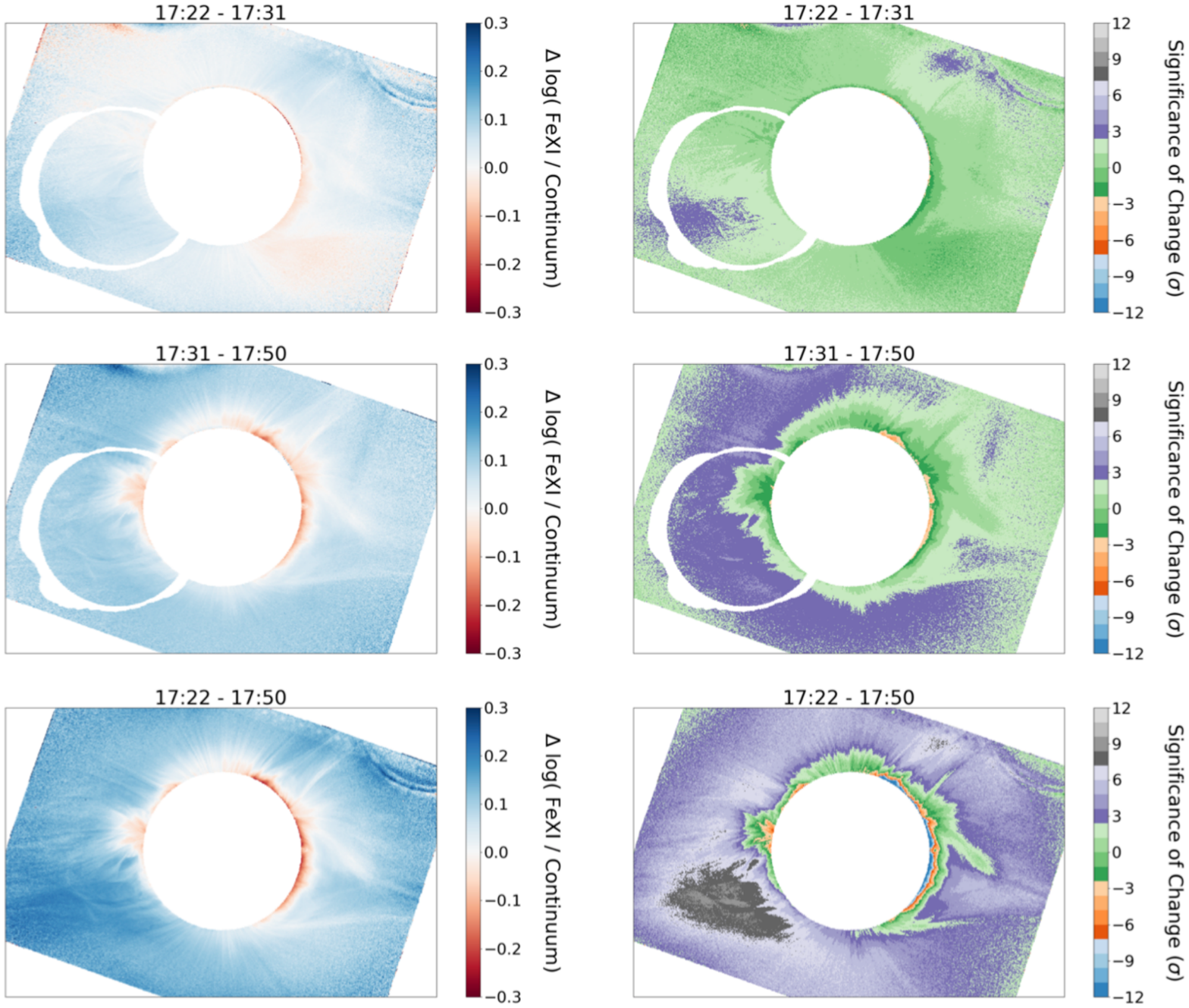}
\caption{Left: Change in the $Log_{10}$ of the \ion[Fe xi] line to continuum ratio between the three observing sites (see Section \ref{lineCont} and \ref{time}). Right: Statistical significance of these changes between sites. Top: Mackay - Mitchell; Middle: Alliance - Mackay; Bottom: Alliance - Mitchell. The white rings in the top and middle panels are the same indication of a ghost image as in Figure \ref{fig5}.}
\label{fig7}
\end{figure*}

Observations of the \ion[Fe xi] and \ion[Fe xiv] line emission (see Section \ref{eclipse}) were calibrated relative to continuum emission observations (see Section \ref{Calib}) enabling the measurement of ionic emission normalized by the continuum emission from 1 to 3 \Rs, as shown in Figure \ref{fig5}. Identical observations were made at three separate observing sites, enabling the quantification of changes of the line to continuum intensity ratio over the 28 minutes between the time of totality at the first and last site as shown in Figures \ref{fig6} and \ref{fig7}. We only display regions of the maps which had a signal to noise ratio $>5$ (from photometric and calibration uncertainties, see Section \ref{Calib}) in the narrowband images used for each plot.

%%FIGURE 4

\par
The dynamics of the line to continuum observations are complex to analyze as they will change as a result of temperature, density and velocity effects. Since we are observing the line emission with very narrowband filters ($\approx 0.5$ nm width), any plasma with a radial velocity (RV) faster than $\approx 190$ km s$^{-1}$ for \ion[Fe xi] and $\approx 280$ km s$^{-1}$ for \ion[Fe xiv] will be Doppler shifted out of our line emission observations (i.e. shifted out of the filter bandpass). The continuum emission from the K+F corona will not change due to such a velocity shift, since it is re-emitting a broad continuum spectrum from the photosphere. So the continuum intensity is integrated along a large 3 dimensional column along the line of sight in contrast to the line emission observation, which will only observe coronal plasma near the plane of sky with little to no Doppler velocity. The line to continuum measurements are therefore indicating the total line emission of particles with no RV relative to the global line of sight corona continuum. 
\par
As shown in Figure \ref{fig5}, the ionic emission relative to the continuum spans about 3 orders of magnitude in the corona for \ion[Fe xiv], while \ion[Fe xi] emission is confined to less than a single order of magnitude of intensity variation in the corona. Streamer regions have the highest \ion[Fe xiv] emission, typically staying around the same intensity as the continuum. In the coronal holes, \ion[Fe xiv] is far less abundant, going down to around 1 $\%$ that of the continuum. \ion[Fe xi] emission is stronger than the continuum emission throughout the entire corona, and is strongest in the western limb active region (see Section \ref{spacecraft}) where it is as much as six times the continuum intensity (note that the scale in Fig \ref{fig5} is limited to provide contrast for lower intensity regions). The strong emission of \ion[Fe xi] around the active region supports the idea of open field regions in the vicinity of active regions, as previously suggested by \cite{Sakao2007} and \cite{Harra2008}.

\par

The low intensity of \ion[Fe xiv] relative to the local continuum is quite a bit lower than what is predicted by \cite{Bemporad2017}, who argue that a single observation of the \ion[Fe xiv] emission without a continuum subtraction can give useful results. We are finding that in fact, the `spectral purity' of \ion[Fe xiv] is less than $50 \%$ in our 0.5 nm bandpasses almost everywhere in the corona, which is more similar to the spectral purity that \cite{Bemporad2017} predicts for a 1 or 2 nm bandpass. Observation and subtraction of the continuum subtraction is absolutely essential to infer any physical properties of the corona other than morphological structure (see Section \ref{eclipse}). Inferring physical properties with an on-band \ion[Fe xiv] data alone is inherently biased and should not be trusted.

\par

%FIGURE 5
\par

%FIGURE 6
\par

The time variations between observing sites are shown for \ion[Fe xiv] (Figure \ref{fig6}) and \ion[Fe xi] (Figure \ref{fig7}) with time changes of 9 (site 1 - 2), 19 (site 2 - 3) and 28 (site 1 - 3) minutes respectively. Statistically significant changes for both ions occurred between the sites, as demonstrated by the right panels in both figures, which indicate the statistical significance of the changes based on the error from both line to continuum measurements. 
\par
The general behavior of the line to continuum ratio indicates that both the line of sight temperature and density distributions were changing globally in the corona. These changes are likely caused by the CME events that occurred prior to the eclipse, as described in Sections \ref{spacecraft}, and \ref{time}. During the 9 minute period between Mitchell, OR, and Mackay, ID, the \ion[Fe xiv] line to continuum ratio is increasing in the low corona, and decreasing in and around the coronal holes. Over the next 19 minutes from Mackay, ID, to Alliance, NE, the \ion[Fe xiv] emission decreases at the site of the CME while increasing slightly in the outer coronal holes. The net effect over the entire 28 minutes between Mitchell, OR and Alliance, NE is a large decrease in \ion[Fe xiv] emission in the coronal holes with a modest increase (still highly statistically significant) in the streamers below $\approx 2$ \Rs. The \ion[Fe xi] line to continuum ratio was growing throughout the corona, with the largest increase centered on the location of the western CME (see Section \ref{spacecraft}). It is important to note that the Mackay and Alliance \ion[Fe xi] line to continuum ratio had to be normalized to the pristine Mitchell data in the north polar hole ($< 1.1$ \Rs) due to some calibration issues (see Section \ref{Calib}), so the increase is relative to the line to continuum ratio in the north polar hole. If line of sight effects from the CME also increased \ion[Fe xi] emission around the coronal hole, then the changes between sites would be even larger than what we have reported. 

\subsection{Electron Temperature Inference}
\label{methods}
The line to continuum observational metric presented in Section \ref{lineCont} is useful for inferring some physical properties of the coronal plasma. However, it is difficult to interpret since it can be altered by density, temperature and velocity effects. To investigate the thermodynamic properties of the solar corona, we used all \ion[Fe xi] and \ion[Fe xiv] imaging data to infer the relative abundance of Fe$^{10+}$ and Fe$^{13+}$ throughout the corona and to calculate an electron temperature corresponding to the ionic abundance ratio. What follows is a description of the method used to make this inference.

\subsubsection{Emission Rules}
\label{emissionRules}

Spontaneous line emission causes the transition of an atom from an excited state (with density $n_u$) to a lower state (with density $n_l$). The rate of emission per unit time can be written as: 
\begin{eqnarray}
\label{Aul}
A_{ul} \ n_u,
\end{eqnarray}
where $A_{ul}$ is the Einstein coefficient for radiation from a given excited state of the atom. 

\par
Similar to emission, the rate of photon absorption is described by:
\begin{eqnarray}
\label{Blu}
B_{lu} \ n_l \ \rho(\nu), 
\end{eqnarray}
where $B_{lu}$ is the Einstein coefficient for absorption and $\rho(\nu)$ is the energy density at the same frequency as the given atomic transition (units of $J \ m^{-3} \ Hz^{-1}$). The relationship between the Einstein coefficients is given by:
\begin{eqnarray}
\label{BtoA}
A_{ul} = B_{lu} \frac{g_u}{g_l} \frac{8 \pi h \nu^3}{c^3},
\end{eqnarray}
with $g_u$ and $g_l$ being the statistical weights of the excited and lower states respectively \citep{Herzberg1945}.
\par
Since we are interested in the total abundance of an ion regardless of its excitation state, we must account for the ion's electronic state distribution. In thermodynamic equilibrium this can be done simply with the Boltzmann formula, but it is not applicable in the low density environment of the corona. In the absence of collisional effects, the level distribution of an ion will depend on the relative rates of radiative absorption spontaneous decay alone (i.e. Equations \ref{Aul} and \ref{Blu}). 

\par
It is possible that non-equilibrium collisional effects could change the level population, but these effects should be unimportant everywhere except in the very low corona. \cite{Habbal2007} showed that the \ion[Fe xi] and \ion[Fe xiv] emission lines specifically will be excited much more frequently by radiative absorption than by collisions, even for densities as high as $n_e = 2 \times 10^7$ cm$^{-3}$. In the very low corona (below $\approx 1.2$ \Rs), collisional effects may become comparable with radiative effects and slightly bias our inferred ionic density ratio if the density is higher than $\approx 10^7$ cm$^{-3}$. Our results should therefore only be considered robust for distances larger than about 1.2 \Rs \ (given the density profile for a coronal hole from \citealt{Cranmer2007}). Streamers will have a higher density, so the distance where collisions are important could be farther away from the Sun in this case. Nevertheless, collisions will increase the emission of $both$ \ion[Fe xi] and \ion[Fe xiv] at a rate proportional to the density squared, and so the line emission ratio should not change substantially due to collisions. 

\par
If there is a high enough energy density of incoming photons, it is possible that some fraction of the ionic density may become trapped at a higher electronic state which would affect the value of $n_u, n_l$, and other possible electronic states. What follows is a short proof that the rate of spontaneous decay is in fact far greater than that of radiative absorption for any arbitrary choice of coronal ionic emission line.
\subsubsection{Level Population Proof}
\label{proof}
Measurements from the International Space Station (ISS) show that the solar spectral energy density at 1 AU has a maximum of about $2 \ J \ s^{-1} \ m^{-2} \ nm^{-1} $ at approximately 550 nm (see the top panel in Figure \ref{fig8}, \citealt{ISS2018}), corresponding to about $2 \times 10^{-12} J \ s^{-1} \ m^{-2} \ Hz^{-1}$. Treating the Sun as a point source results in an intensity of solar photons in the corona (say at the photosphere $R = $\Rs) of $\approx 10^{-7} J \ s^{-1} \ m^{-2} \ Hz^{-1}$. Converting this to an energy density (dividing by the speed of light) gives $\rho(\nu) \approx 3 \times 10^{-16} J \ m^{-3} \ Hz^{-1}$ for the photon energy density in the corona at 550 nm. Multiplying $\rho(\nu)$ by $B_{12}$ (substituting with equation \ref{BtoA}) gives:

%FIGURE 7
\begin{figure*}[t!]
\centering
\includegraphics[width = 4in]{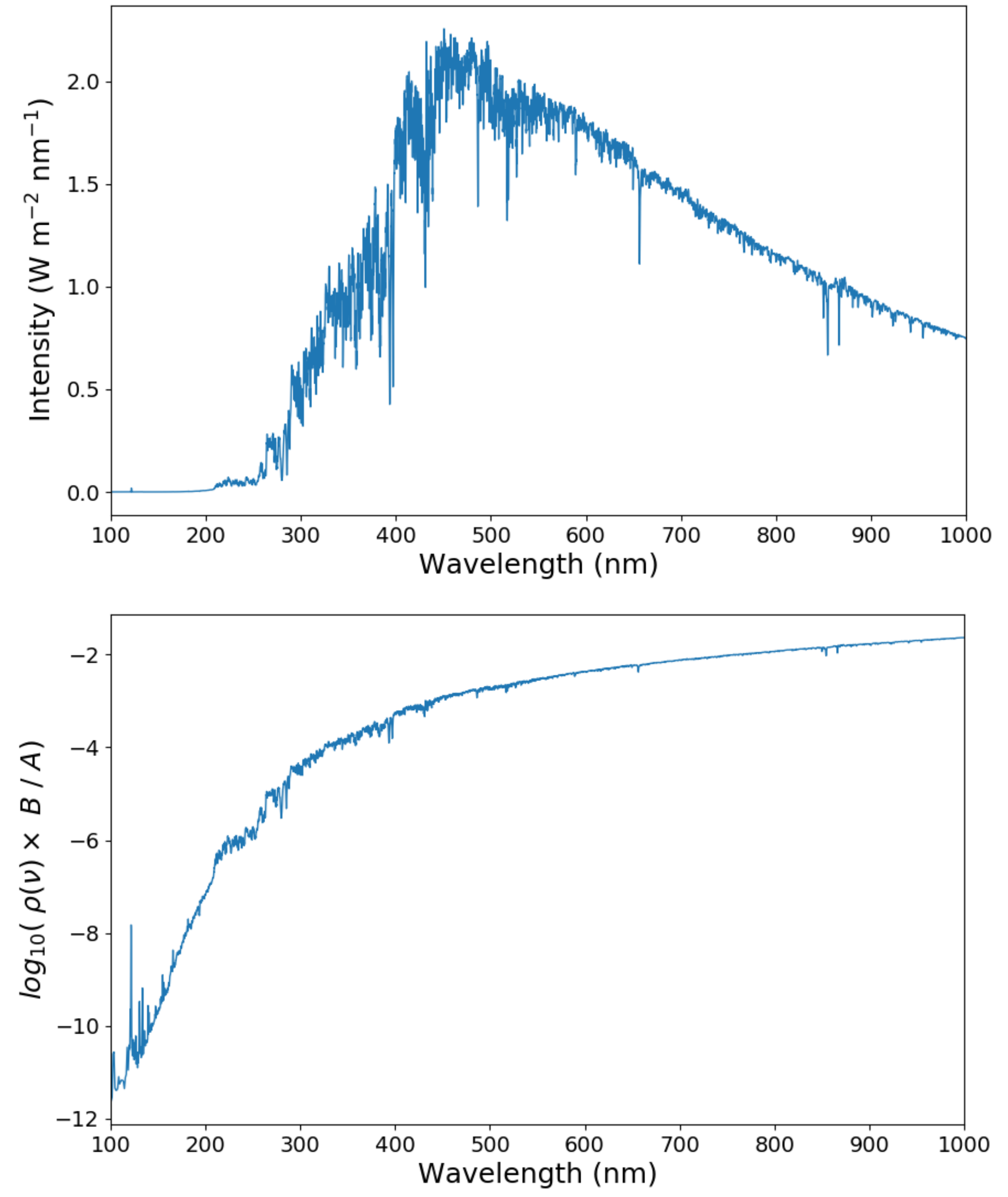}
\caption{Top: Solar spectral energy density from ISS observations at 1 AU \citep{ISS2018}. Bottom: Relative rate of photo excitation and spontaneous emission versus wavelength for any energy level transition based on the energy density from the top curve (scaled to 1 \Rs) as shown in the example in Section \ref{proof}. For all visible (and shorter) wavelengths, the rate of spontaneous emission is at least 2 orders of magnitude smaller than photo-excitation.}
\label{fig8}
\end{figure*}

\begin{eqnarray}
\label{eqnProof}
\rho(\nu) \times B_{lu} = \ 3 \times 10^{-16} \ A_{ul} \ \frac{g_u}{g_l} \frac{c^3}{8 \pi h \nu^3} \ \approx  0.003  \times A_{ul},
\end{eqnarray} 
provided that the statistical weights are the same within an order of magnitude (since the g's are known for a given line, this proof applies to any arbitrary line). Equation \ref{eqnProof} indicates that the rate of spontaneous decay is about $10^3$ times higher than the rate of resonance excitation for ions in the corona at around 550 nm. This same calculation was repeated for the entire solar spectrum in the vicinity of visible wavelengths, as shown in Figure \ref{fig8}. These calculations indicate that the rate of resonant excitation from photospheric photons will be at least two orders of magnitude lower than the rate of spontaneous decay of the same state for every possible electronic level transition with wavelengths ranging from the EUV to the near Infrared. The radiation density provided by the solar photosphere is simply not high enough to cause resonance excitation at a fast enough rate to substantially populate any state other than the lowest level.

\subsubsection{Electron Temperature}
\label{temp}
The example from Section \ref{proof} (see Figure \ref{fig8}) demonstrates that for typical coronal conditions (i.e. $n_e < \approx 10^7$ cm$^{-3}$), one can assume that all ions are in the ground electronic state, except for rare instances when they absorb photons and immediately decay spontaneously. Thus $A_{ul} >> B_{lu} \times \rho(\nu)$, and the density of the excited state, $n_u$, will be quickly depleted to the ground state, $n_l$, for all possible energy states. This depletion of higher states will cause the total density of the ion, $n_i$, to be about equal to the density in the ground state ($n_{i} \approx n_l$). Therefore, every emitted photon from the coronal ions will originate from a very recent photo-absorption, and the total number of photons emitted per unit time will be given by equation \ref{Blu} for any coronal ion emission line that ends in the ground state. 
\par

The intensity in our eclipse images (see Section \ref{eclipse}) for the on-band, $I_{on,i}$, and the continuum $I_{cont,i}$, can be related by:
\begin{eqnarray}
\label{eqnOn}
I_{on,i} - I_{cont,i} = B_{lu, i} \ \rho(\nu_{i}) \ n_{i} \ \epsilon_{i},
\end{eqnarray}  
where $\epsilon_{i}$ is the efficiency of the photometric observations (after calibration, see Section \ref{Calib}). 
\par

Now consider two different ionization states $j$ and $k$ (analogous to Fe$^{10+}$ and Fe$^{13+}$) and take the ratio of equation \ref{eqnOn} for $j$ and $k$,\begin{eqnarray}
\label{eqnOn2}
\frac{I_{on,j} - I_{cont,j}}{I_{on,k} - I_{cont,k}} = \frac{B_{j} \ \rho(\nu_{j}) \ n_{j} \ \epsilon_{j}}{B_{k} \ \rho(\nu_{k}) \ n_{k} \ \epsilon_{k}}.
\end{eqnarray} 
Substituting $B_{j}$ and $B_{k}$ from equation \ref{BtoA} gives,
\begin{eqnarray}
\label{eqnOn3}
\frac{I_{on,j} - I_{cont,j}}{I_{on,k} - I_{cont,k}} = \frac{\rho(\nu_{j}) \ n_{j} \ \epsilon_{j} \ A_{j} \ g_{u,j} \ g_{l,k} \ (\nu_j)^{-3}}{\rho(\nu_{k}) \ n_{k} \ \epsilon_{k} \ A_{k}  \ g_{l,j} \ g_{u,k} \ (\nu_k)^{-3}}. 
\end{eqnarray}
\par

Finally solving for the density ratio gives:
\begin{eqnarray}
\label{eqnFinal}
\frac{n_{j}}{n_{k}} = \frac{(I_{on,j} - I_{cont,j}) \ \rho(\nu_{k}) \ \epsilon_{k} \ A_{k}  \ g_{l,j} \ g_{u,k} \ {\nu_{k}}^3}{(I_{on,k} - I_{cont,k}) \ \rho(\nu_{j}) \ \epsilon_{j} \ A_{j} \ g_{u,j} \ g_{l,k} \ {\nu_{j}}^3},
\end{eqnarray}
where $\frac{\rho(\nu_{k}) \ \epsilon_{k}}{\rho(\nu_{j}) \ \epsilon_{j}}$ is measured as the continuum ratio in the low corona between the continuum data sets for each line after the on- and off-band pairs were self-calibrated with solar disk observations (see Section \ref{Calib} for details).

\par
\begin{figure*}[t!]
\centering
\includegraphics[width = 6.75in]{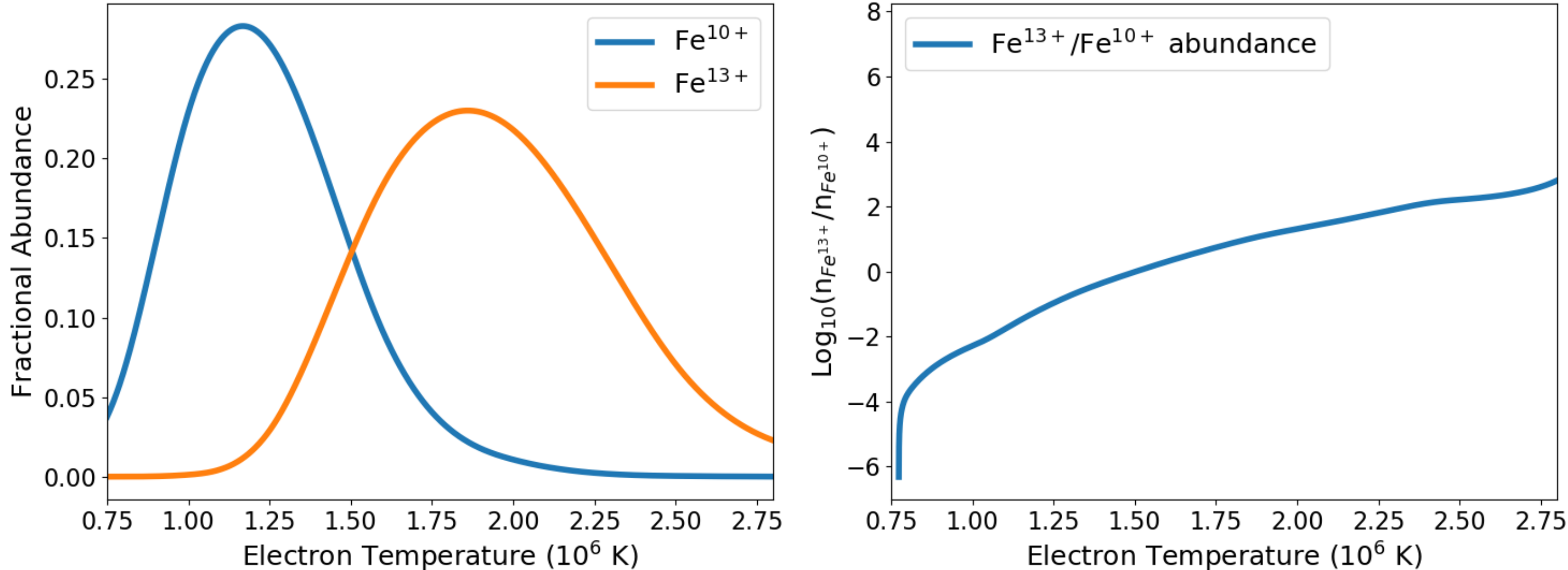}
\caption{Temperature dependence of the fractional ionic abundance of Fe$^{10^+}$ and Fe$^{13^+}$ interpolated from the calculations of \cite{Arnaud1992} (left) and of the abundance ratio (right).}
\label{fig9}
\end{figure*}

Equation \ref{eqnFinal} provides a method for using four different imaging observations, namely a continuum observation combined with the on-band observations of two different optical forbidden emission lines of the same element (with a relative calibration), to directly calculate the relative density of two ions. We computed equation \ref{eqnFinal} separately for every line of sight to measure the relative density of Fe$^{10+}$ and Fe$^{13+}$ throughout the corona at each site. The constants used in this calculation are shown in Table \ref{table2}. The relative density maps are then used to infer the electron temperature based on the results of \cite{Arnaud1992}, who calculated the ionization equilibrium abundances as a function of $T_e$ for many states of Fe (see Figure \ref{fig9}). The \cite{Arnaud1992} abundance values were calculated in steps of $\Delta Log(T_e) = 0.1$; we have interpolated their data for this work. 

\begin{table}[h]
\begin{center}
\begin{tabular}{ c  c  c  c  c c }
\hline
Line &  $\lambda_{ion}$ (nm) & $\nu_{ion}$ ($10^{14}$ Hz)  & $A_{ion}$ ($s^{-1}$) & $g_l$ & $g_u$  \\
\hline
 \ion[Fe xi]  & 789.2 & 3.801 &43.7 & 5 & 3  \\
\hline
 \ion[Fe xiv]  &530.3 & 5.657 & 60.2 & 2 & 4   \\
\caption{Constants used for equation \ref{eqnFinal}. Data from NIST (\citealt{NIST} and ref. therein)}
\label{table2}
\end{tabular} 
\end{center}
\vspace{-9mm}
\end{table}

\par
\cite{Arnaud1992} used a low density approximation where ``the steady state ionic fractions do not depend on electron density," which is the same assumption that we are making in our analysis. The \cite{Arnaud1992} results also assume ionization equilibrium, which is not necessarily valid beyond $\approx 1.5$ \Rs \ (see Section \ref{intro}; \citealt{Landi2012b}). We will not consider non-equilibrium effects here, but rather report on the inferred $T_e$ based on the relative ion density alone (inferred from emission assuming no collisional effects), which is then directly comparable with inferred coronal $T_e$ values from \textit{in situ} ionic charge state measurements. We chose to use the \cite{Arnaud1992} abundances due to the lack of dependence on free parameters. A more complex model to measure the true kinetic electron temperature would require constraints on the density and outflow velocity that were not possible to measure directly with our data and would induce additional uncertainty into the results. Our observations are also not measuring $T_e$ for a single isothermal unit of plasma, but rather we are measuring the line of sight average emission of \ion[Fe xi] and \ion[Fe xiv]. Our inferred $T_e$ values are thus providing a density weighted average of the electron temperature distribution near the plane of sky.
\par
It is also important to remember that once the plasma reaches the freeze-in distance $\approx 1.2 - 1.4$ \Rs \ for coronal holes and up to $1.6 - 1.8$ \Rs \ for streamers, as shown by \cite{Boe2018}, the ionic abundance (and so the inferred $T_e$) will remain constant as the plasma flows outward in the solar wind. Beyond the freeze-in distance, our inferred $T_e$ will be representative of what the $T_e$ was in the low corona when the plasma was below the freeze-in distance. In fact, the ionic abundances may decouple from $T_e$ even below the freeze-in distance as shown by \cite{Landi2012b}.

%FIGURE 8
\par

\par
\begin{figure*}[t!]
\centering
\includegraphics[width = 6in]{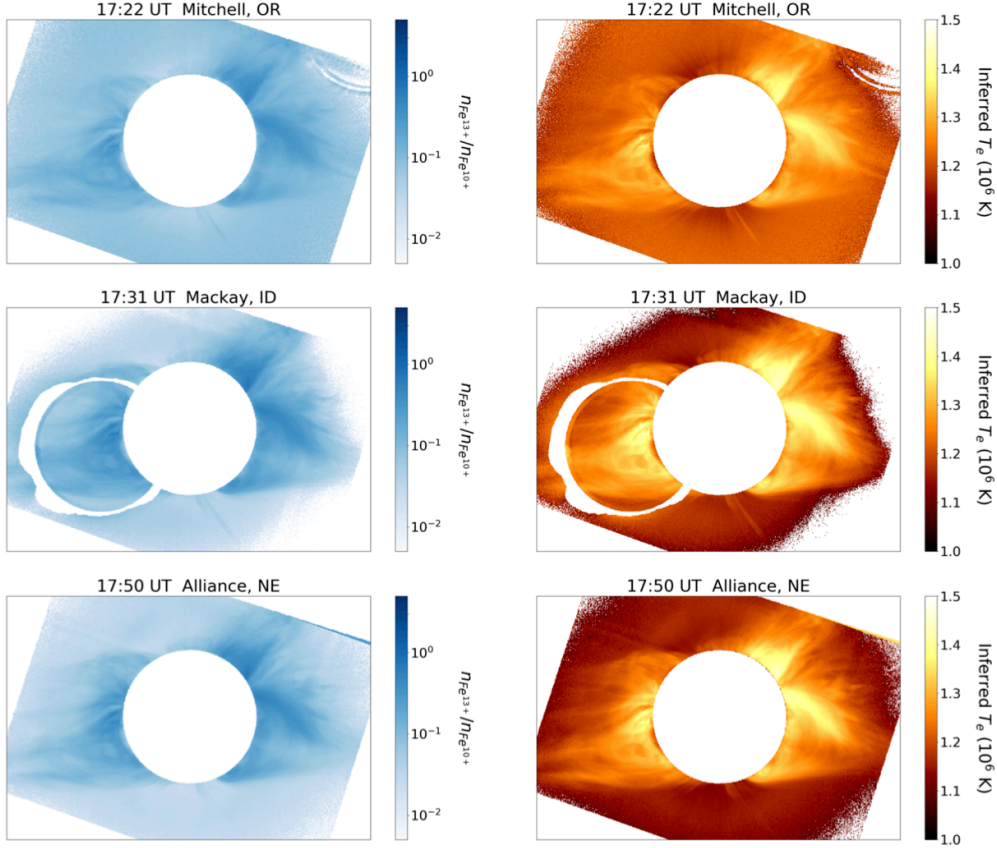}
\caption{Left: Density ratio of Fe$^{13^+}$ to Fe$^{10^+}$ as calculated using equation \ref{eqnFinal} (see Section \ref{temp}) from the observations of \ion[Fe xi] and \ion[Fe xiv] made at three separate sites. Right: Inferred electron temperature ($T_e$) from the relative density measurements that were fed through the temperature function shown in Figure \ref{fig9}. The white ring in the middle panels indicates the location of a ghost image as in Figure \ref{fig5}.
}
\label{fig10}
\end{figure*}

The inferred Fe$^{13+}$ to Fe$^{10+}$ ionic density ratio and corresponding $T_e$ are shown in Figure \ref{fig10}. The density ratio maps show that the coronal holes at the north and south poles have a Fe$^{13+}$ density that is about 2 to 5 $\%$ of the Fe$^{10+}$ density. On the other hand, the streamers contain a much larger amount of Fe$^{13+}$, which ranges from 10 to 50 $\%$ of the Fe$^{10+}$ density. Overall, the ionic density ratio varies by over an order of magnitude in the corona. The inferred $T_e$ in coronal holes is the lowest, about 1.1 to 1.2 $\times 10^6$ K, whereas streamers have the highest temperatures, ranging from 1.2 to 1.4 $\times 10^6$ K. The coronal regions which have the highest $T_e$ correlate well with the presence of closed magnetic field lines, underlying prominences (as shown before by \citealt{Habbal2010b}), and a high continuum intensity (i.e. high electron density). 
\par

%FIGURE 9

\begin{figure*}[t!]
\centering
\includegraphics[width = 6in]{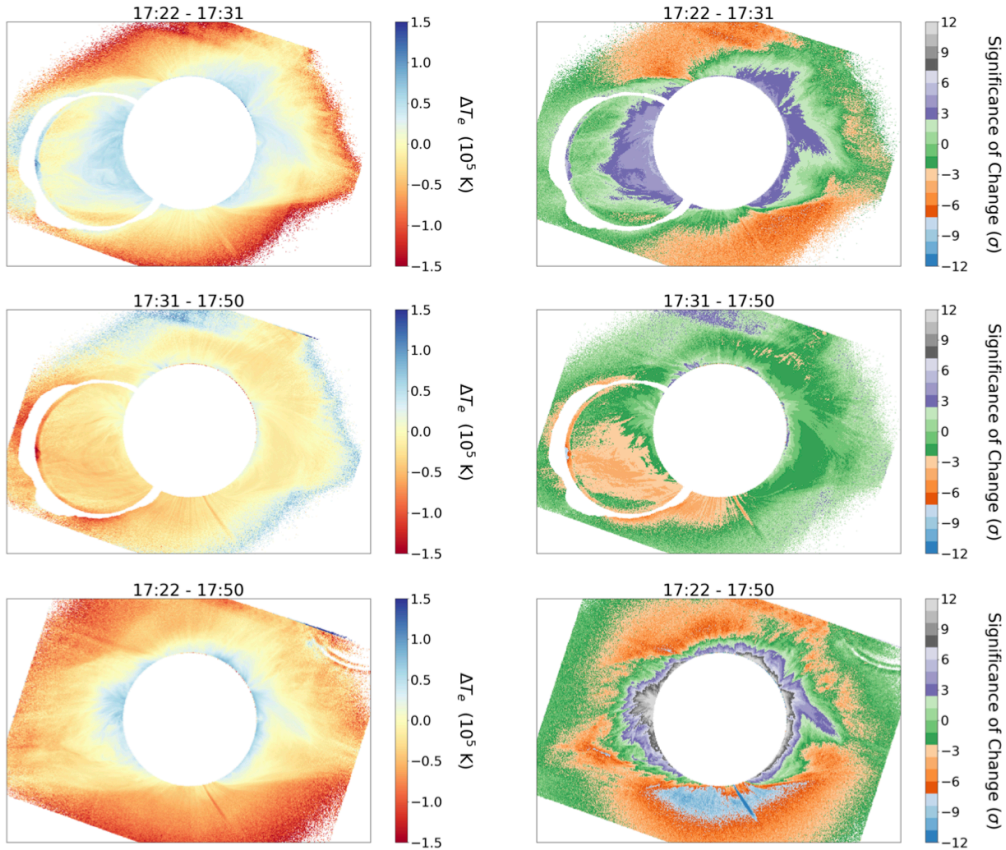}
\caption{Changes in the inferred $T_e$ (left) between each of the three sites (see Section \ref{temp}). Top: Mackay - Mitchell; Middle: Alliance - Mackay; Bottom: Alliance - Mitchell. The white rings in the top and middle panels are the same indication of a ghost image as in Figure \ref{fig5}.}
\label{fig11}
\end{figure*}

The coronal $T_e$ changed significantly along most lines of sight between the sites as shown in Figure \ref{fig11}, with the largest change occurring between the first and last site. The outer regions of the corona are found to be cooling with time while the inner regions are heating up. Many of these changes are significant to greater than 3 $\sigma$, with the lower regions of streamers commonly changing by $> 10 \sigma$ based on propagated calibration and $\sqrt{N}$ photometric errors (where $N$ is the number of data counts in the original images, see Section \ref{eclipse}). A discussion on the nature of these changes is given in detail in Section \ref{time}.

%FIGURE 10
\par

%\newpage

\section{Discussion}
\label{disc}
\subsection{Spatial Distribution of $T_e$}
\label{tempDist}

Inferences of the coronal electron temperature ($T_e$) throughout the corona (see Figure \ref{fig10}) indicate that each individual field line has a unique temperature structure. The large variability in $T_e$ in streamer regions implies the existence of a preponderance of structures in the corona, each with unique temperature and density profiles depending on the exact physical parameters which act upon the field line. A physically realistic coronal heating model should be able to explain the diversity of temperature conditions present along various field lines. 
\par
The coronal holes have inferred $T_e$'s between 1.1 and 1.2 $\times 10^6$ K, which is in agreement with previous UV and EUV emission measurements that yielded coronal hole temperatures in the range of $1.1-1.3 \times 10^6$ K \citep{Habbal1993, Wit2013}, albeit in the very low corona ($<1.2$ \Rs). The same temperature range for polar coronal holes was also inferred at large heliocentric distances using charge state measurements over multiple solar cycles by the Ulysses and ACE spacecraft \citep{Smith2003, Habbal2010a}. The consistency of coronal hole temperatures inferred, using vastly different techniques across multiple solar cycles, hints that coronal heating mechanisms within open magnetic field line regions are relatively stable throughout multiple cycles. 
\par

The inferred $T_e$ values in streamers and closed field line regions range from 1.2 to 1.4 $\times 10^6$ K. These $T_e$ values are consistent with the quiet sun regions inferred by SDO/AIA below 1.2 \Rs, but are far lower than the typical active region temperatures of $2-4 \times 10^6$ K (e.g. \citealt{Wit2013, Morgan2017}). This difference is likely caused by observational selection effects that are very different in optical versus EUV observations. The solar photosphere emits very little in the EUV (see Figure \ref{fig8}), so the rate of radiative excitation for EUV lines is much smaller than for optical lines. Coronal EUV emission is consequently dominant where it is collisionally excited ($\propto n_e^2$) in the low corona and closed field line regions ($<$ 1.2 \Rs), while optical emission lines will originate from radiatively excited plasma ($\propto n_e$) on both open and closed field lines near the plane of sky out to at least $2 -3$ \Rs. It is reasonable to expect that observations of different sets of emission lines that operate with different non-equilibrium excitation mechanisms will result in a different temperature inference. Regardless, our metric for inferring the electron temperature is a self-consistent one that is useful for inferring dynamic and thermodynamic changes in the corona, even if it is not an exact kinetic electron temperature for a single isothermal plasma element. \par

\par
Another key difference between the work presented here and $T_e$ studies using SDO/AIA data is the temperature response of the observed emission lines. We used emission from Fe$^{10+}$ and Fe$^{13+}$, which will be sensitive to plasma with temperatures between about 1 and 2 $\times 10^6$ K (see Figure \ref{fig9}). Any plasma at a higher or lower temperature than that range will not be visible in our observations. SDO/AIA has bandpasses which are sensitive to $T_e$ as high as $10^7$ K, and so they are capable of measuring higher $T_e$ plasma (e.g. \citealt{ODwyer2010, Boerner2012, Morgan2017}). Data \textit{in situ} from Ulysses and ACE over multiple solar cycles indicates that the average coronal $T_e$ is around $1.2-1.5 \times 10^6$ K \citep{McComas1998, Gloeckler1998, Smith2003}, while there still are occasional short duration spikes in $T_e$ that can go as high as $10^7$ K (e.g. \citealt{Habbal2010a}). Combining \textit{in situ} and emission line observations leads to the interpretation that the vast majority of the plasma in the corona is $< 1.5 \times 10^6$ K except for some higher $T_e$ plasma originating from closed field lines in and around active regions. Some of this hotter plasma can be released via CMEs and reconnection events into the solar wind, but it will only compose a small fraction of the total solar wind plasma. 

\subsection{Thermodynamic Temporal Variability}
\label{time}
As noted earlier in Section \ref{lineCont}, there were significant changes in the \ion[Fe xi] and \ion[Fe xiv] emission relative to the continuum (see Figs. \ref{fig5}, \ref{fig6}, and \ref{fig7}), and in the Fe$^{13+}$ to Fe$^{10+}$ ionic density ratio, and thus $T_e$ between our three eclipse observing sites (see Figs. \ref{fig10} and \ref{fig11} and Section \ref{temp}). Changes in the inferred $T_e$ are significant to $> 3 \sigma$ throughout a large portion of the corona (based on photometric and calibration uncertainties, see Section \ref{Calib}), with some regions changing as much as $> 10 \sigma$. In general we find that around streamers the corona is increasing in temperature by up to $0.7 \times 10^5$ K, while the outer corona is cooling down by as much as $1.0 \times 10^5$ K. The same general behavior is present over every time difference, with the largest changes corresponding to the longest time period between the first and last site.

 \par
 
Since a radial velocity of $\approx 190$ km s$^{-1}$ for \ion[Fe xi] and $\approx 280$ km s$^{-1}$ for \ion[Fe xiv] will shift the line emission out of our narrowband observations (see Section \ref{lineCont}), the line to continuum ratio and $T_e$ metrics will be highly sensitive to velocity perturbations in the corona. The inferred density and temperature changes do not necessarily require that specific plasma change its kinetic electron temperature, but rather can be explained if the corona is disturbed enough to change the average plasma distribution along the line of sight with a Doppler velocity less than about 200 km s$^{-1}$.  Additionally, a radial outflow speed of only 200 km s$^{-1}$ would result in a travel distance of 0.5 \Rs \ over the half hour from Mitchell, OR to Alliance, NE. So it is possible for much of the corona to substantially change its density and temperature distribution along a single line of sight during that time, since we are not really observing the exact same plasma at each site. 

\par

The most likely cause of the inferred line emission and $T_e$ changes is the set of eruptions that took place just prior to the time of the eclipse (see Section \ref{spacecraft}). CMEs have previously been shown to have secondary `sympathetic CMEs' that commonly occur elsewhere in the corona in the first few hours after an initial eruption event \citep{Moon2003}. Sympathetic CMEs are thought to be caused by magnetically connected regions near the solar surface, such as filaments and active regions, where a CME in one region can stimulate connected regions to erupt. The dynamics of the small eruptive events of the two western streamers can be explained in this manner. It is likely that they would not have erupted on their own, and indeed they were stable enough to not generate a CME. Instead the perturbation from the CME in the SE, combined with the erupting active region near disk center, caused them to undergo a minor sub-CME (i.e. failed CME) scale eruption. At the time of the Mitchell, OR, the corona was observed in the period immediately following these eruptions. Therefore, it is likely that we have observed the corona recovering from the eruptive perturbations.

\par
The observations at Mitchell occurred 4 hours after the beginning of the CME activity. Consequently, the Mitchell data do not truly represent the initial conditions of the corona without a CME present. It is thus somewhat difficult to interpret the exact physical cause of the line to continuum ratio and $T_e$ changes, as the changes are relative to an already perturbed state. One possible interpretation is that a CME shockwave passed through the corona prior to our first observation. Consequently, the abundance of  \ion[Fe xiv] may have been enhanced in the Mitchell data. The decrease of \ion[Fe xiv] in the coronal holes would then represent the departure of this plasma from the corona (or at least accelerated to a Doppler velocity $> 200$ km km s$^{-1}$). An increasing \ion[Fe xiv] in the lower regions of the streamers could then be explained as either heating from the CME activity, or by the magnetic field lines in the streamers repairing themselves by re-closing field lines that were ripped open by the CMEs (which would increase the density of hot plasma). 
\par
Similarly, the \ion[Fe xi] intensity may have been suppressed at the time of Mitchell, so the increase may be due to cold plasma repopulating the plane of sky after the CME disturbed it. The location of the largest increase of \ion[Fe xi] intensity is just below the CME front in the south western corona, which supports this interpretation. However, the \ion[Fe xi] changes are relative to the northern coronal hole (due to calibration issues, see Section \ref{Calib}). If the \ion[Fe xi] intensity was actually dropping in the coronal hole, then the interpretation of the data would be somewhat different. Regardless, we can say with confidence that there were dramatic changes in the corona on short time scales, and that CME activity is the driving cause. 

\par

Even though the CME and sub-CME eruptive events that occurred prior to the eclipse were not strikingly large in either LASCO or AIA data, these events caused significant changes to the thermodynamics of coronal plasma on a scale of only 10 to 30 minutes. Small scale CMEs have been found previously to have a substantial impact on the structure and brightness of field lines in the corona using total solar eclipse white light data (similar to Fig. \ref{fig1} A, \citealt{Alzate2017}), but never before have CMEs been shown to have such a global impact on the electron temperature and density within the corona below 3 \Rs. Future modeling efforts should strive to explain the cause of such rapid CME-driven changes in the mid corona. 
\par

While such changes have never been reported directly in the corona, CMEs have been shown to have substantial impacts in the heliosphere. Studies using \textit{in situ} detectors have long found shock waves resulting from CMEs in interplanetary space (e.g. \citealt{Gosling1968, Hundhausen1970}). Ionic state measurements have indicated that CME shock fronts do increase the ionization state of solar wind particles, and so will increase the inferred coronal $T_e$ of such events (e.g. \citealt{Bame1979, Fenimore1980}). The interpretation of the dynamics in the corona has been limited though, due to the difficulty in robustly tracing back \textit{in situ} measurements to precise sources in the corona \citep{Galvin1997}. 
\par

Direct observations in the corona during total solar eclipses, as presented here, are currently the best technique to probe the temperature dynamics in the solar corona between 1 and 3 \Rs. Even the new generation of ground-based solar telescopes such as DKIST will not be able to measure such rapid and global changes in the corona as we have presented in this study, due to their limited spatial extent.

\section{Conclusions}
\label{conc}
In this work, we have presented observations of the \ion[Fe xi] and \ion[Fe xiv] optical forbidden line emission between 1 and 3 \Rs \ from three different sites taken during the 21 August 2017 total solar eclipse (see Section \ref{eclipse}). The line to continuum ratio (Fig. \ref{fig5}) was determined for each ion at each site and used to infer the relative density of Fe$^{10+}$ and Fe$^{13+}$ (Fig. \ref{fig10}) and compute a coronal electron temperature ($T_e$) via theoretical abundances (as described in Section \ref{temp}). We find that:
\begin{enumerate}
\item $T_e$ in the corona ranges from 1.1 to 1.4 $\times 10^6$ K. Open field line structures such as coronal holes have the lowest temperatures ($1.1-1.2 \times 10^6$ K) while streamers have the hottest temperatures ($1.2-1.4 \times 10^6$ K). These results are consistent with previously published inferences (see references in Section \ref{disc}).
\item Statistically significant temporal changes of the line to continuum intensity ratio of \ion[Fe xiv] (Fig. \ref{fig6}) and \ion[Fe xi] (Fig. \ref{fig7}), and of $T_e$ (Fig. \ref{fig11}) occur throughout the corona over the 28 minutes between the first and last observing site. The outer corona is decreasing in temperature (up to $10^5$ K), while the inner corona is increasing in temperature (up to $0.7 \times 10^5$). Almost the entire corona had the inferred $T_e$ change by $> 3 \sigma$, with a sizable fraction changing by $> 10 \sigma$. These changes are likely due to the impact of a CME and two smaller eruptions just prior to the eclipse (see Sections \ref{spacecraft}, \ref{time}). 
\item When observing emission lines at visible wavelengths, it is critical to measure the corresponding continuum radiation to subtract electron scattering (i.e. K corona) and scattering by interplanetary dust (i.e. F corona) to correctly isolate ionic emission (see Section \ref{eclipse}). Even in a small bandpass, the continuum radiation constitutes a substantial fraction of the emission at optical wavelengths. We thus caution the `spectral purity' arguments of \cite{Bemporad2017}, who state that correction for the continuum is not necessary. Indeed, using an on-band observation alone will not produce physically meaningful results (especially for \ion[Fe xiv], see Section \ref{lineCont}).
\end{enumerate}

\par
This work highlights the value of multi-site narrowband imaging for inferring $T_e$ via charge state ratios throughout the corona, and its temporal evolution. If ground- or space-based coronagraph telescopes are equipped with a similar suite of filters, as used in this work, then inferences of $T_e$ and the impact of dynamic events such as CMEs could be studied with high cadence and over long periods of time with the same techniques demonstrated here.

%%%%%%%%%%%%%%%%%%%%%%%%%%%%%%%%%%%%%%%%%%%%%%%%%%%%%%%%%%%%%%%%
%
%  ACKNOWLEDGMENTS
\subsection*{Acknowledgments}
Special thanks go to Judd Johnson for designing and building the narrowband camera systems used in this work. We thank Martina Arndt, Garry Nitta, Marcel B{\'e}l{\'i}k and Radovan Mrll{\'a}k for their assistance in operating the narrowband camera systems during totality. Thanks also to Nathalia Alzate and Huw Morgan for sharing the processed SDO/AIA and SOHO/LASCO C2 images (in Fig \ref{fig2}). Financial support was provided by NASA grant NNX17AH69G and NSF grants AGS-1358239, AGS-1255894, and AST-1733542 to the Institute for Astronomy of the University of Hawaii. AAS provided partial support to S. Habbal with grant ORS001869, and AURA/NSO supported B. Boe under grant N97991C to the Institute for Astronomy at the University of Hawaii. M. Druckm\"uller was supported by the Grant Agency of Brno University of Technology, project No. FSI-S-14-2290. 

\appendix

\section{Doppler Shifted Line Emission}
\label{DopplerShift}

A blue shift Doppler velocity of $\approx 530$ km s$^{-1}$ can contribute additional Doppler shifted ionic emission in the continuum bandpass used at Mackay and Alliance. We can expect possible emission from the \ion[Fe xi] line at that velocity, since Fe$^{10+}$ is highly abundant in the fast solar wind ($> \approx 600$ km s$^{-1}$, e.g. \citealt{Smith2003, Habbal2010a}). The Fe$^{13+}$ ion is much more abundant in the slow solar wind (between 200 and 600 km s$^{-1}$), so \ion[Fe xiv] emission is possible but not as likely in large quantities at a velocity of 530 km s$^{-1}$. 
\par
\begin{figure*}[h!]
\centering
\includegraphics[width = 6in]{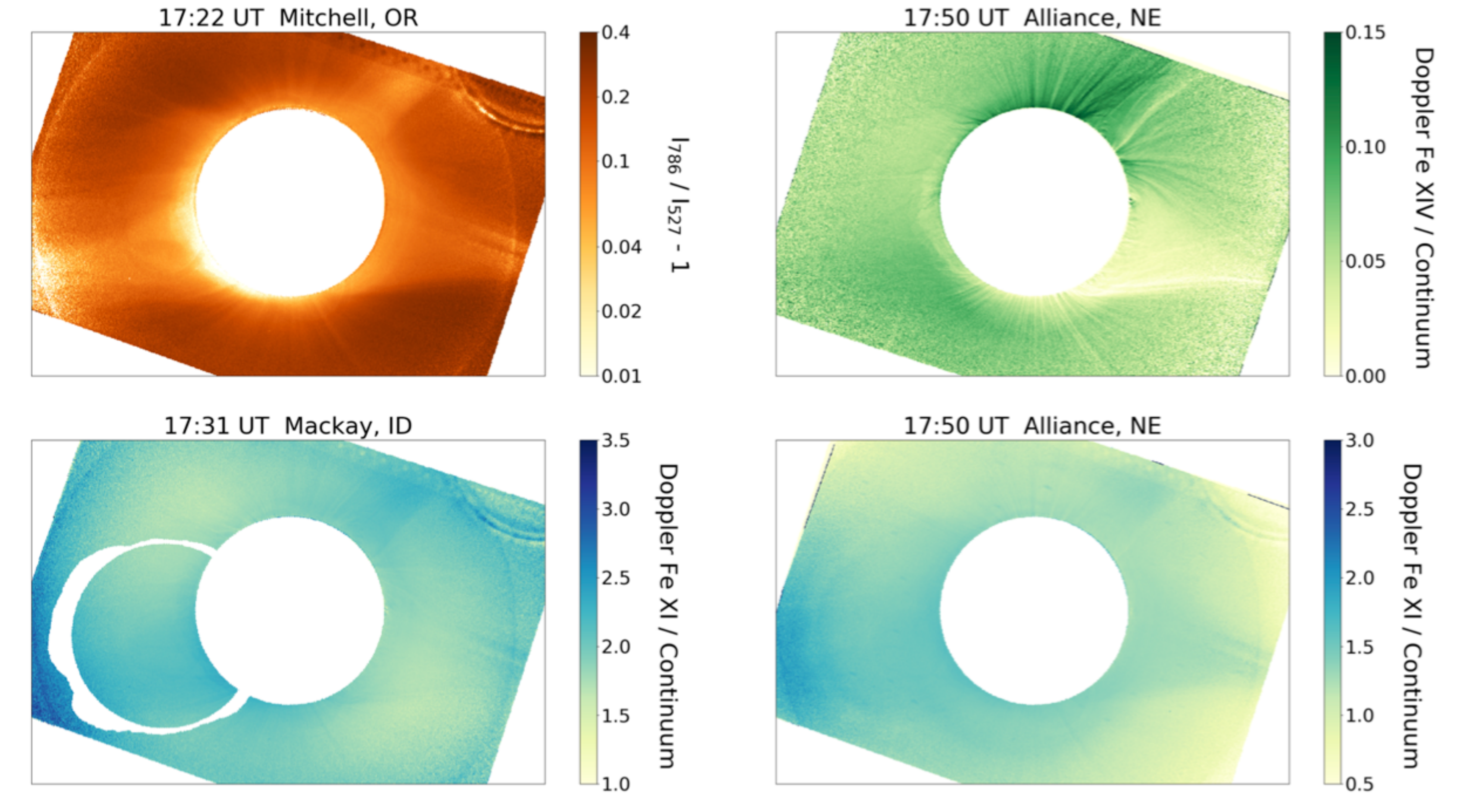}
\caption{Top left: Excess of emission at 786 nm compared to 527 nm at Mitchell. Top Right: Doppler shifted ($\approx 530$ km s$^{-1}$) line to continuum ratio of \ion[Fe xiv] measured at Alliance. Bottom Left: Doppler shifted line to continuum ratio of \ion[Fe xi] from Mackay. Bottom Right: Same as left but for Alliance. The line and continuum data were calibrated using the technique outlined in Section \ref{Calib}. The white ring in the middle panels indicates the location of a ghost image in the raw data produced by internal reflections in the optics (see Appendix \ref{ghost}).}
\label{figA}
\end{figure*}

Thankfully, the Mitchell data are pristine (see Section \ref{Calib}), and we had an additional 552.3 nm continuum observation which we used as a direct substitute for the \ion[Fe xiv] continuum at the Alliance site. To utilize `green' continuum observations for the \ion[Fe xi] 789.2 nm line, we first had to account for any reddening in the corona due to dust scattering of the F corona (see Section \ref{eclipse}). We measured the coronal color by taking the continuum intensity ratio for the Mitchell 527.4 nm and 786.1 nm data, after a relative calibration in the western streamer below 1.1 \Rs (where the K corona dominates). The resulting reddening map is presented in Figure \ref{figA}. The Mackay and Alliance \ion[Fe xi] continuum data were then generated by using green continuum data with an enhancement from the F corona reddening map (Refer to Section \ref{Calib} for details on the calibrations and generation of the synthetic continuum data for Mackay and Alliance).  
\par

%%FIGURE A

Once new continuum data had been created for the line emission observations at Mackay and Alliance, we were able to measure the quantity of Doppler emission in the original off-band data. Removing the 552.3 nm continuum from the Alliance data indicates a small amount of Doppler shifted \ion[Fe xiv] emission, with an average intensity of $5.7\%$ relative to the continuum (see Figure \ref{figA}). The location of the largest Doppler shifted \ion[Fe xiv] intensity matches the location of the decreasing \ion[Fe xiv] emission in the plane of sky coronal holes (see Section \ref{lineCont}). The largest Doppler emission in the north coronal hole could also be caused by the halo-CME originating from the central disk active region, which was at a latitude of about 10 degrees north. A halo-CME from that active region would likely be pointed slightly northward and would explain the structure of the Doppler shifted \ion[Fe xiv] observation. Given the small amount of Doppler shifted \ion[Fe xiv] emission: We used the possibly contaminated \ion[Fe xiv] continuum at Mackay, but added an extra 5.7$\%$ uncertainty to the Mackay data. 
\par
The \ion[Fe xi] Doppler emission had a much stronger intensity than \ion[Fe xiv] (see Fig \ref{figA}), with some lines of sight having a Doppler emission intensity of 3 times that of the continuum. Given the 0.5 nm width of the bandpasses, the Doppler observations will observe emission contributions from everything between about 350 and 700 km s$^{-1}$, which explains why the Doppler emission can still be stronger than the continuum. The Mackay data has a higher \ion[Fe xi] Doppler intensity throughout (about a 10 to 30 $\%$ difference), but both Mackay and Alliance have the same basic structure of the highest emission occurring at the location of the western CME (see Figure \ref{figA}). Since the CME is emitting \ion[Fe xi] at 530 km s$^{-1}$, the velocity vector of the CME has to be somewhat out of the plane of sky.  Such a vector would also explain why the CME was not reported as a major event by LASCO coronagraph data (see Section \ref{spacecraft}). Nonetheless, continuum intensity variation in the LASCO C2 data around the time of the eclipse does support halo-CME activity (see Fig \ref{fig4}). The overall Doppler \ion[Fe xi] intensity decrease between Mackay and Alliance hints that either the CME was passing through this velocity space temporarily (i.e. the CME may have been accelerating), or that the ionic emission is decreasing as the CME expands. The ACE spacecraft also saw a CME signature three days after the eclipse that supports the presence of a halo-CME (see Section \ref{spacecraft}). The CME arrival time delay would correspond to a constant velocity of about 545 km s$^{-1}$ if the CME began at 13:00 UT on the day of the eclipse, which is consistent with the observed Doppler shifted emission. 
\section{Ghost Image Removal}
\label{ghost}
In some of the narrowband imaging cameras, internal reflections from the optics resulted in the appearance of secondary `ghost' images of the corona on the detector. The ghost images are significantly fainter, typically $\leq 2 \%$ of the original intensity. Fortunately, the intensity of the corona drops exponentially with distance from the Sun, so only the inner portion of the reflected corona has any substantial intensity in the ghost images. Unfortunately, there will be a small systematic bias due to the ghost image between $\approx 1-1.5 $ \Rs. 
\par
To correct for the bias caused by the ghost image, we removed the signature of ghost images by modeling the reflection and subtracting it. Given that the ghost is a reflection of the original image, we subtracted the image from itself with a physical offset ($\delta X, \delta Y$) and an intensity multiplier to simulate the brightness of the ghost ($B_g$). We also accounted for the ghost image being out of focus by smoothing the original image with a Gaussian filter using $\sigma = \sigma_g$ and rotated the ghost frame by $180^{\circ}$ before subtraction, to account for the rotation of the mirrored image caused by the telescope's internal optics (a double mirror reflection in the narrowband filter). We first set the parameters manually then allowed perturbations in an automated Monte Carlo brute force technique to improve the ghost subtraction. 

\par
\begin{figure*}[h!]
\centering
\includegraphics[width = 4.5in]{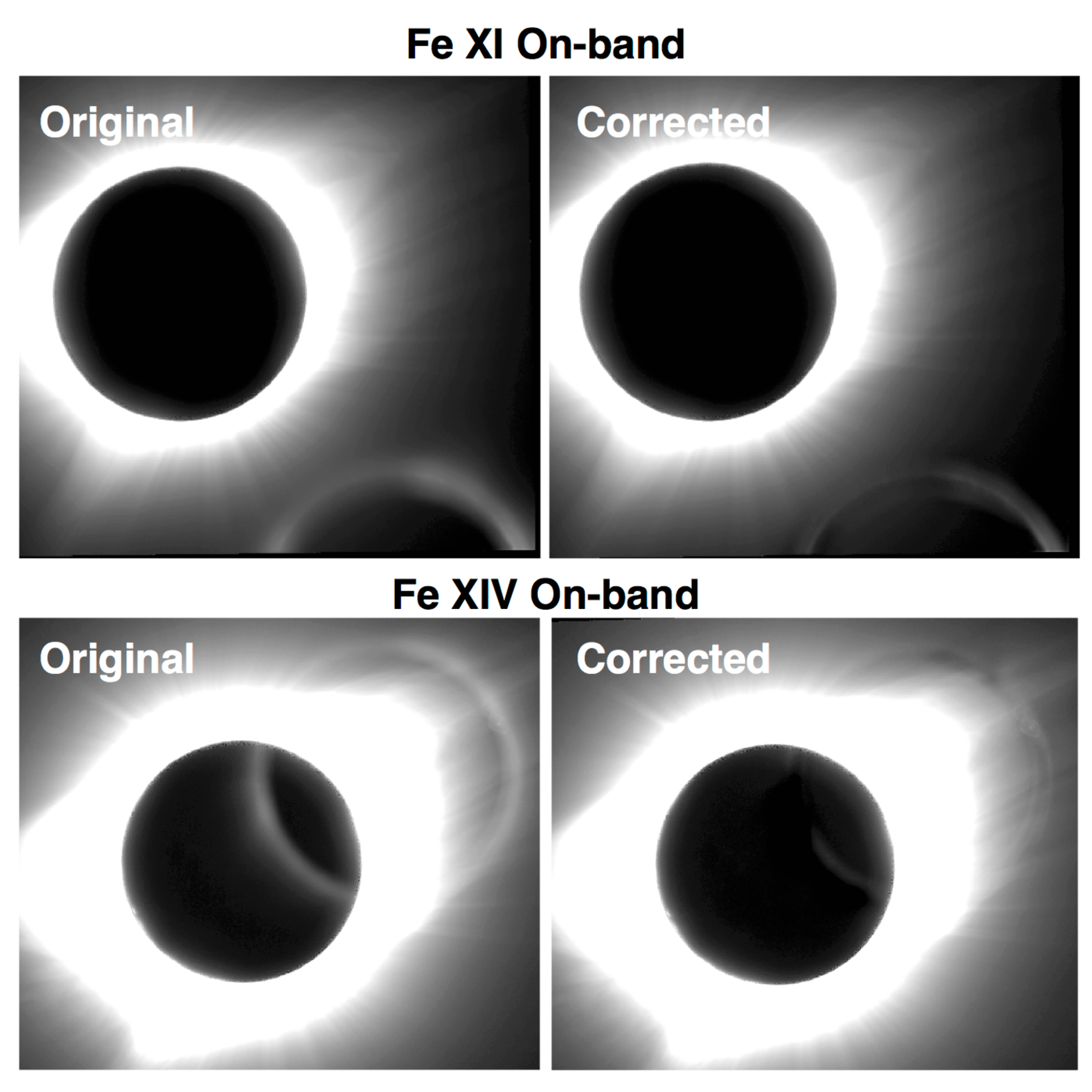}
\caption{Example of the ghost image removal on \ion[Fe xi] (top) and \ion[Fe xiv]  (bottom) on-band data (not off-band subtracted) from the Mackay, ID site. The left column shows a cropped version of the original images with logarithmic scaling and somewhat saturated in order to show the ghost images clearly. The final ghost subtracted images are shown on the right with scaling identical to the original images.}
\label{figB}
\end{figure*}

Samples of the ghost removal results are shown in Figure \ref{figB}, with frames from before and after the subtraction. This technique is not perfect, often leaving visible traces of the ghost image especially in a ring that corresponds to the reflection of the bright inner corona. The benefit of doing this subtraction method is to remove systematic bias that the ghost image has outside of the obvious ghost ring. 

%FIGURE B
\par

\par
We successfully performed this removal process to ghost images in the far corners of the Mitchell \ion[Fe xiv] off-band, Mackay \ion[Fe xi] on-band and Alliance \ion[Fe xiv] on-band images which had no visible effect on our results. Ghost images were also removed from the Alliance \ion[Fe xi] off-band and Mackay \ion[Fe xiv] on-band images, which had more centrally located ghost reflections. The Alliance \ion[Fe xiv] ghost was very faint and could be removed almost completely, leaving only a very slight sign of the ring. Unfortunately, the inner ring of the ghost image in the Mackay \ion[Fe xiv] on-band data (shown in Figure \ref{figB}) still had a large amount of contamination due to the brightness of the ghost image. We masked the region of the Mackay \ion[Fe xiv] ghost image where the ghost ring had substantially contaminated the image in order to reduce the chances of misinterpretation that the ghost ring is some real physical structure. The masked ring can be seen in the panels involving Mackay data, specifically Figures \ref{fig5}, \ref{fig6}, \ref{fig7}, \ref{fig10}, \ref{fig11}, and \ref{figA}.

\bibliographystyle{apj}
\bibliography{Boe2019ArXiv}

\end{document}